\documentclass[twocolumn,superscriptaddress,bibnotes,secnumarabic,aps,longbibliography,floatfix]{revtex4-2}

\usepackage{footmisc}
\usepackage{amsmath}
\usepackage{amssymb}
\usepackage{amsfonts}
\usepackage{color}
\usepackage{graphicx}
\usepackage{lipsum,graphicx}
\usepackage{amsfonts}
\usepackage{textcomp}
\usepackage{hyperref}
\usepackage{notes2bib}
\usepackage{natbib}

\usepackage{footnote}

\newcommand{\beq}{\begin{equation}}
\newcommand{\eeq}{\end{equation}}
\newcommand{\bea}{\begin{eqnarray}}
\newcommand{\eea}{\end{eqnarray}}

\begin{document}

\title{Universal negative magnetoresistance in antiferromagnetic metals caused by symmetry breaking of electron wave functions}
\author{Pavel D. Grigoriev}
\affiliation{L.D. Landau Institute of Theoretical Physics, RAS, Chernogolovka, 143423, Russia}
\affiliation{National University of Science and Technology ''MISiS'', 119049, Moscow, Russia} 
\affiliation{HSE University, Moscow 101000, Russia}
\author{Nikita S. Pavlov}
\affiliation{Institute for Electrophysics,  RAS, Ekaterinburg, 620016, Russia}
\affiliation{V.~L.~Ginzburg Research Center at P.~N.~Lebedev Physical Institute, RAS, Moscow 119991, Russia}
\author{Igor A. Nekrasov}
\affiliation{Institute for Electrophysics, RAS, Ekaterinburg, 620016, Russia}
\author{Igor R. Shein}
\affiliation{Institute of Solid State Chemistry, RAS, Ekaterinburg, 620990, Russia}
\author{Andrey V. Sadakov}
\author{Oleg A. Sobolevskiy}
\affiliation{V.~L.~Ginzburg Research Center at P.~N.~Lebedev Physical Institute, RAS, Moscow 119991, Russia}
\author{Evgeny  Maltsev}
\affiliation{Leibniz Institute for Solid State and Materials Research, IFW Dresden, 
	D-01069 Dresden, Germany}
\affiliation{Dresden-W\"{u}rzburg Cluster of Excellence ct.qmat, Dresden, Germany}
\author{Vladimir M. Pudalov}
\affiliation{V.~L.~Ginzburg Research Center at P.~N.~Lebedev Physical Institute, RAS,  Moscow 119991, Russia}
\date{\today}

\begin{abstract}
Layered van der Waals crystals of topologically non-trivial and trivial semimetals with antiferromagnetic (AFM)
ordering of magnetic sublattice are known to exhibit a negative
magnetoresistance that is well correlated with AFM magnetization changes in
a magnetic field. This effect is reported in several experimental studies 
with EuFe$_2$As$_2$, EuSn$_2$As$_2$,  EuSn$_2$P$_2$, etc., where the resistance decreases quadratically with
field by about $5\%$ 
	up to the spin-polarization field.
Although this effect is well documented experimentally, its theoretical
explanation is missing up to date. Here, we propose a
theoretical mechanism describing the observed magnetoresistance that
is inherent in AFM metals and is based on violation the binary $\hat{T}_{2}$ symmetry.
It is almost isotropic to the field and current directions, contrary to the known mechanisms such as giant magnetoresistance and chiral anomaly.
The proposed intrinsic mechanism of magnetoresistance is
strong in a wide class of the layered AFM-ordered semimetals. 
The theoretically calculated magnetoresistance is qualitatively consistent with experimental
data for crystals of various composition.
\end{abstract}

\maketitle

Topology incorporated with magnetism provides a fertile playground for studying novel quantum states 
and  therefore attracts tremendous attention. 	Topological aspects and related electronic phenomena  
occuring in antiferromagnetic (AFM) materials are among the central topics in condensed matter physics. Clarifying the underlying mechanisms that govern mutual effects of the magnetic ordering, broken symmetry,  exchange and spin-orbit coupling  on magnetotransport in  layered crystals is crucial for comprehending emergent properties and phenomena in spintronics. 
	
The vast majority of  layered AFM crystals of topological insulators (TI) and  semimetals   exhibit  such 
	characteristic feature as a negative magnetoresistance (NMR) in  external magnetic field. 
	This effect has recently attracted  a great deal of interest, 	as it is 	related with chiral anomaly in Dirac  and Weyl semimetals  \cite{li_NatCom_2015, yan_PRR_2022}, with  Chern insulators and  anomalous quantized Hall state in AFM topological insulators  	\cite{li_PRB_2019, ge_NatlSciRev_2020}.
Indeed, for topological semimetals (Na$_3$Bi, TaS, Cd$_3$As$_2$), a large NMR results from the chiral anomaly  \cite{Xiong, HuangTaAsNMR2015} and appears  when electric field is applied nearly parallel to magnetic field. 
	This  NMR is highly anisotropic and its sign changes as the angle between the magnetic field $\boldsymbol{H}$ and current 
	increases to $\pi/2$ \cite{Xiong}. In the topological insulator MnBi$_2$Te$_4$, 
	a huge NMR \cite{ge_NatlSciRev_2020}  occurs as a result of the  topological phase transition 
	 from  AFM TI to a ferromagnetic Weyl semimetal in external field.

On the other hand, a number of   layered  crystals of topologically trivial AFM semimetals 
		(EuSn$_2$As$_2$ \cite{arguilla_InChemFront_2017, chen_ChPhysLet_2020, 
			li_PRB_2021},  EuSn$_2$P$_2$  \cite{gui_ACSCentSci_2019},   CaCo$_2$As$_2$ \cite{ying_PRB_2012}, CsCo$_2$Se$_2$ \cite{yang_JMMM_2019}, etc.)  exhibit  {\em isotropic}  negative magnetoresistance 
		(NIMR) for transport both along and across the layers. 
 There is no  	satisfactory treatment of such isotropic NMR. Particularly, the magnetoresistance (MR) related with magnon scattering is negative only in ferromagnets and arises from the magnetic-field-induced suppression of scattering  \cite{fontes_PRB_1999}.
For antiferromagnets, the magnon scattering causes {\em positive} MR \cite{yamada-takada_1973-1,yamada-takada_1973-2} 
\footnote{We presume that for  our case of the easy magnetization $ab$ plane, the  $\mathbf{H}\|ab$ and $\mathbf{H}\|c$ field orientations correspond, in notations of Ref.~\cite{yamada-takada_1973-2}, to the  easy axis  $\mathbf{H}\|z$, and perpendicular to it $\mathbf{H}\|x$,  respectively.} 
that originates from a field-induced increase in spin fluctuations.

Negative giant MR (GMR) is well known for the Fe/Cr superlattices where resistance decreases as magnetization in the neighbouring Fe-layers turns with external  magnetic	field  from antiparallel to parallel \cite{baibich_PRL(1988), dagotto_PhysRep_2001}. For ideal interfaces,  GMR  should develop for charge transport across the layers \cite{dagotto_PhysRep_2001}. In experiments with artificially deposited  Fe/Cr layers \cite{baibich_PRL(1988)}, a weaker NMR is observed also for current along the layers and is associated with diffuse scattering at layer boundaries  \cite{campley_PRL_1989, barthelemy_enciclopedia}.
Another case, the ``colossal'' NMR  is known for doped manganites and magnetic granular samples \cite{dagotto_PhysRep_2001, yin_PRB_2000, barthelemy_enciclopedia} where domains due to phase separation effects and misoriented crystallites lead to electron  scattering at the  boundaries. 
	
	 In case of high quality {\em single crystals}, the scattering by grains, domain or  crystallite boundaries is missing;  also,  there is not much sense to consider  roughness of atomic layers.
The isotropy of experimentally observed NIMR encourages us to consider scattering by point defects.

	In this paper we suggest a 
	mechanism of the {\em intrinsic} isotropic NMR  that is irrelevant to magnons, magnetic impurities, grain boundaries, layer roughness, 
	and topology, being general to AFM metals and rather strong in a wide class of the {\em anisotropic} layered crystals of AFM metals and semimetals. 
These materials exhibit  a negative magnetoresistance that is tightly correlated with the magnetization field dependence 
\cite{jiang_NJP_2009, chen_ChPhysLet_2020, li_PRB_2021, sanchez_PRB_2021, gui_ACSCentSci_2019}.
The  magnetization changes linearly with external field and  sharply  saturates above a  field of complete spin polarization  $H_{\rm sf}$  (see {\bf Supplementary Note 1}); the $H_{\rm sf}$ field is often called spin-flip field for the easy axis case. Though the close relationship between magnetization and magnetoresistance is known also for
GMR and colossal MR (CMR)  \cite{dagotto_PhysRep_2001, yin_PRB_2000}, the suggested mechanism is completely different.

\begin{figure}[tbp]
\includegraphics[width=0.47\textwidth]{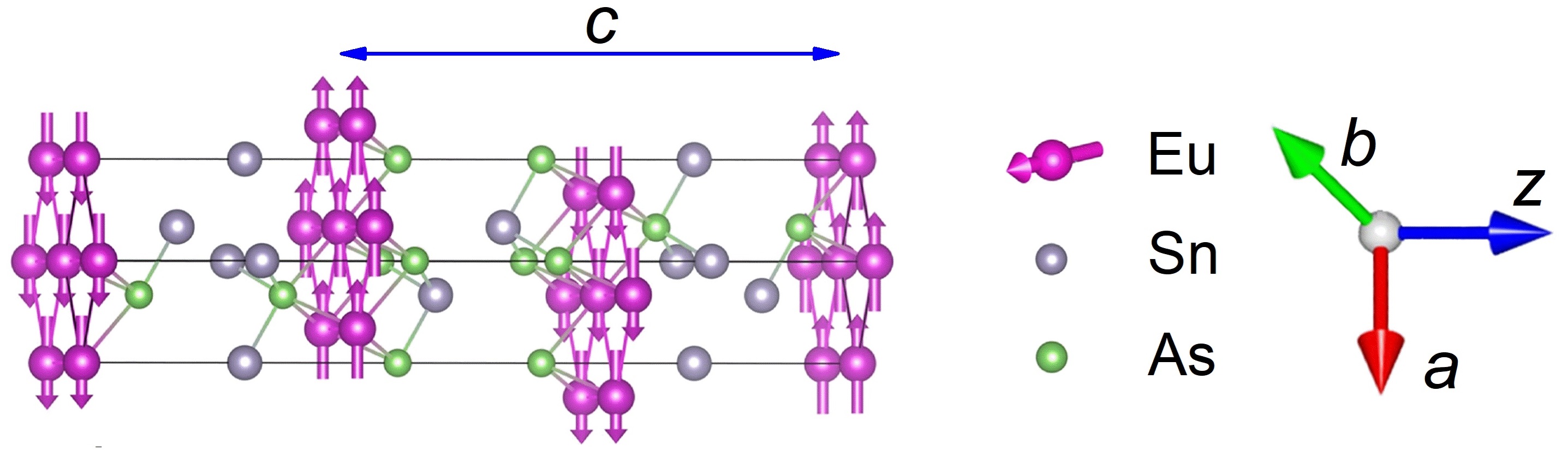}  
	\caption{{\bf Schematic picture of the EuSn$_2$As$_2$ lattice structure (adapted from Ref.~\cite{golov_JMMM_2022})}.
		Magenta arrows show Eu-atoms magnetization direction in the A-type AFM
		ordered state. For more detail on the Eu magnetic moments orientation, see Supplementary Note 3. Horizontal arrow  denotes the lattice spacing $c\approx 2.64$nm along $z$-axis   \cite{arguilla_InChemFront_2017, chen_ChPhysLet_2020, PRB_tbp}. 
	}
	\label{fig:1}
\end{figure}

 We test our model by 	comparing it with experimentally measured magnetoresistance in  EuSn$_2$As$_2$
  that has  a topologically trivial BS (see {\bf Supplementary Note 2}). 
For  this representative  compound, at $T<24$K the magnetic Eu-sublattice experiences magnetic  
ordering into the A-type AFM structure where Eu magnetic moments lie in the easy $ab$-plane and
rotate by $\pi$ from layer to layer along the $z$-axis (see Fig.~\ref{fig:1}a, and also {\bf Suplementary Note 3}). 

Synchronous with  magnetization, the diagonal resistivity decreases approximately parabolically with field as $\delta\rho_{xx}(H) \propto -\alpha H^2$. 
At higher fields $H>H_{\rm sf}$,  magnetotransport perpendicular to the field sharply changes from NMR to a conventional positive magnetoresistance.  Such NMR in layered AFM crystals, closely correlated with the field dependence of magnetization $M(H)$ \cite{jiang_NJP_2009, chen_ChPhysLet_2020, li_PRB_2021, gui_ACSCentSci_2019}, 
was discussed previously in terms of either nontrivial topological properties, or in terms
of scattering by domains, grain boundaries etc., with no detailed theoretical consideration.

 In this paper we address the issue of the origin of negative isotropic MR. Specifically,  
 	we (i) propose a model where NIMR universally originates from the enhancement of electron scattering in AFM crystals,   
	(ii) substantiate this proposal using the density functional theory (DFT) calculations of spin polarized charge distribution, 
		and (iii) compare the model quantitatively with our experimental data for NIMR in EuSn$_2$As$_2$ and qualitatively -- with other  compounds of the same class of layered AFM crystals.\\
		
{\Large\bf Results}\\

\begin{figure}[tbh]
\includegraphics[width=0.5\textwidth]{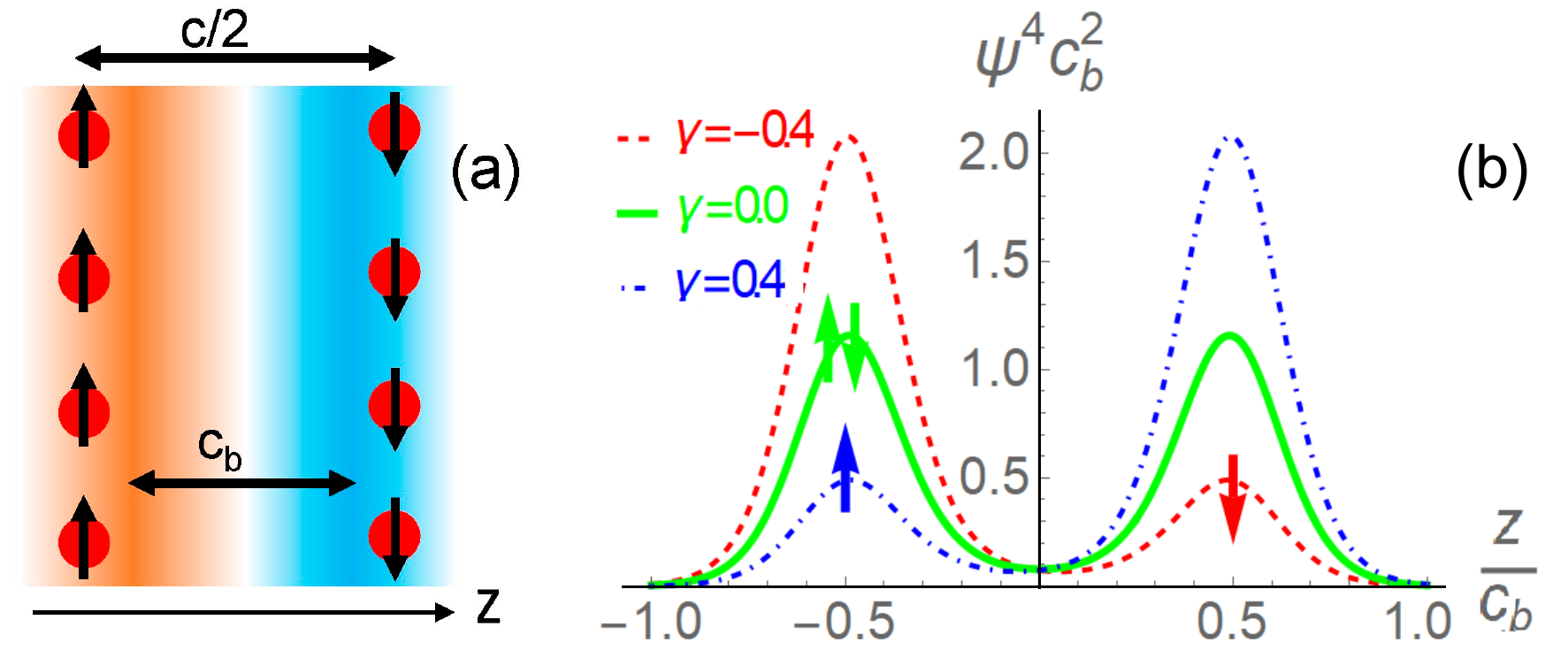}
\caption{{\bf Wave function (WF) magnitude distribution along the crystal $z$-axis 
	for the lowest-energy quantum state [Eq.~(\protect\ref{psipmWF})] in a
	double-well potential, modeling two AFM sublattices.} 
	{\bf a} Schematic color illustration of the spin-up (red) and spin-down (blue) 
wave function distribution along the $z$-axis in the half-cell of the $c/2$-size. 
$c_b$ denotes the distance between the two spin-split WF maxima.   
Color intensity represents
the WF magnitude. Red circles show Eu atoms, and arrows -- their magnetization direction in the AFM state. 
	{\bf b}  Schematic picture of the fourth power $%
\protect\psi ^{4}(z)$ of electron WF given by Eq. (\ref{psipmWF}), entering the scattering rate in Eq. (\protect\ref%
{tau}). The asymmetry parameter  $\protect\gamma \equiv  E_{ex}/2t_0 =0$
(green solid line), $\protect\gamma=-0.4$ (dashed red line) and $\protect%
\gamma=0.4$ (dash-dotted blue line). 
}
\label{FigWF}
\end{figure}

{\large\bf Model and a qualitative consideration}
\\
The main idea of the proposed NIMR mechanism is as follows. The AFM order violates the binary $\hat{T}_{2}$ symmetry $1\leftrightarrow 2$ between two magnetic sublattices  (adjacent layers in  EuSn$_2$As$_2$, see Fig. 1).
This $\hat{T}_{2}$ symmetry to the permutation of AFM sublattices is irrelevant to topology and means equal energies and weights of electron eigenfunctions on each AFM sublattice. 
 Its violation  leads to spatial redistribution of wave functions (WF) $\psi _{\sigma }\left( \boldsymbol{r}\right) $ for two different spin projections ($\sigma=\uparrow\downarrow$) and to their concentration near the corresponding AFM sublattices. 
 As a result of this WF squeezing,  the electron scattering rate \cite{Abrik} by short-range crystal defects or $\delta$-correlated disorder 
	increases already in the Born approximation:
\begin{equation}
1/\tau \propto \int \left\vert \psi _{\sigma }\left( \boldsymbol{r}\right)
\right\vert ^{4}d^{3}\boldsymbol{r} . \label{tau}
\end{equation}%
 Figure \ref{FigWF} schematically illustrates this $\hat{T}_{2}$-symmetry violation of
 electron eigenfunctions by AFM order and the resulting enhancement of $\left\vert
\psi _{\sigma }\right\vert ^{4}(\boldsymbol{r})$ entering the scattering rate in Eq. (\ref{tau}). 

As we show below  in Eqs. \eqref{psipmWF} and \eqref{dI}, the degree of $\hat{T}_{2}$ symmetry
violation,  given by Eq. \eqref{psipmWF}, 
 and the corresponding $1/\tau $ enhancement,  given by Eq. \eqref{dI}, 
depend on the ratio $
\gamma = E_{ex}/2t_{0}  $
of the exchange splitting $ E_{ex}$ of  conduction electron bands to their
hopping amplitude $t_{0}$ between the opposite AFM sublattices  (Fig.~\ref{fig:1}). In the main order in $\gamma $ this relative enhancement of $1/\tau $ is
\begin{equation}
	-\delta \tau /\tau \approx \gamma^2 = ( E_{ex}/2t_{0})^2 . \label{dtau}
\end{equation}%

While $t_{0}$ is determined by the band structure and does not considerably depend on magnetic field $H$,
the exchange splitting $ E_{ex}\propto L_{AFM}\left( H\right) $
decreases with $H$ according to the mean-field theory (see, e.g. Eq.\,(3) of Ref.~\cite{MandelJETP1973}
and Eq.\,(24) of Ref.~\cite{StreitPRB1980}): 
\begin{equation}
\frac{ E_{ex}(H)}{ E_{ex}(H=0)}=\frac{L_{AFM}(H)}{%
L_{AFM}(0)}
\approx \sqrt{1-\left(\frac{H}{H_{\mathrm{sf}}}\right)^{2}}.  
	\label{EexB}
\end{equation}%
Here $\boldsymbol{L}_{AFM}$ is the AFM order parameter,  i.e. the difference of magnetization $\boldsymbol{M}_{1,2}$ of two AFM sublattices: $\boldsymbol{L}_{AFM}=\boldsymbol{M}_{1}-\boldsymbol{M}_{2}$. Eq. \eqref{EexB} can be derived as follows.
In a sufficiently large field $H\leq H_{\mathrm{sf}}$, the AFM order parameter $\boldsymbol{L}_{AFM}\perp \boldsymbol{H}$ \cite{kittel}. In this range the field also causes a net magnetization $\boldsymbol{M} =\boldsymbol{M}_{1}+\boldsymbol{M}_{2}=\chi \boldsymbol{H}$, coming mainly from the nonzero averaged spins $\langle \boldsymbol{S}\rangle $ of magnetic atoms, Eu in our case. The corresponding magnetic susceptibility $\chi $ is approximately constant as a function of field at temperature $T<T_N$ below the Neel temperature $T_N$ \cite{kittel}: $\chi (H) \approx const$. This is also confirmed by direct AC susceptibility and DC magnetization \cite{pakhira_PRB_2021} in a field $H\leq H_{\mathrm{sf}}$. The total spin of each magnetic ion is fixed, $S_{||}^2+S_{\perp}^2={\hat S}^2=S(S+1)$, where $S_{||}$ and $S_{\perp}$ are its components parallel and perpendicular to magnetic field, giving the net magnetization $\boldsymbol{M}$ and AFM order parameter $\boldsymbol{L}_{AFM}$ correspondingly. Then $L_{AFM}/M=\langle S_{\perp}\rangle /\langle S_{||}\rangle $ leads to 
\begin{equation}
	\boldsymbol{L}_{AFM}^2+\boldsymbol{M}^2=\boldsymbol{L}_{AFM}^2+(\chi\boldsymbol{H})^2=const 
	\label{LM}
\end{equation}%
and to the right side of Eq. \eqref{EexB}.

For our model of negative magnetoresistance, we only need the exchange interaction between the magnetic ions (or even their sublattices) and conduction electrons. It is nearly a point-like Heisenberg interaction $-\cal{J}\boldsymbol{S}\boldsymbol{\sigma}$ between the spin $\boldsymbol{S}$ of magnetic ion and the spin $\boldsymbol{\sigma}/2$ of conduction electrons, because the exchange coupling is strong only where the wave functions of conduction electrons overlap with the localized electrons of magnetic ions. The wave function $\psi_{\sigma}(r)$ of conduction electron is delocalized and extended to many magnetic ions. Therefore, $\psi_{\sigma}(r)$ feels an average spin $\langle \boldsymbol{S}_i \rangle$ of magnetic ions on each AFM sublattice $i$, which is proportional to its total magnetization $\boldsymbol{M}_i$. Hence, instead of considering the interaction between individual Eu atoms and electrons, we only need the mean-field average spin $\langle \boldsymbol{S}_i \rangle \propto \boldsymbol{M}_i$, which creates a spin-dependent potential for conduction electrons.

The spin $\boldsymbol{S}$ of magnetic atoms is only reoriented by magnetic field, leaving $\boldsymbol{L}_{AFM}\perp \boldsymbol{H}$ over the intire range  $H< H_{sf}$. Hence, the absolute value $S=|\boldsymbol{S}|$ of spin moment and of the exchange-splitting of conducting electrons is independent of magnetic field. This splitting from electron exchange interaction is usually calculated using the DFT in collinear form, especially for Eu with half-filled 4f orbitals carrying a local moment. We denote this splitting $ E_{ex0}=E_{ex}(H=0)$ in Eq. \eqref{EexB}; below from our DFT calculations we estimate it to be $E_{ex0}\approx 30$meV for EuSn$_2$As$_2$. The exchange splitting $E_{ex}=E_{ex}(H)$,
entering our NIMR mechanism and Eqs. \eqref{dtau}-\eqref{EexB}, is the difference of exchange energy on two different AFM sublattices. The AFM order parameter $\boldsymbol{L}_{AFM}$ also gives the difference between averaged spin orientation of magnetic atoms on these two AFM sublattices, which determines the exchange splitting $E_{ex}=E_{ex}(H)$. Hence, ${L}_{AFM}(H) \propto E_{ex}(H)$, as we write in the left part of Eq. \eqref{EexB}. 

Indeed, for the NIMR effect, coming from the redistribution of $\psi_{\sigma}(r)$ between two AFM sublattices as given by Eq. \eqref{psipmWF} below, we need the difference of weights of $\psi_{\sigma}(r)$ on two different AFM sublattices. And this difference of weights is proportional to the difference $E_{ex}$, entering the Hamiltonian in Eq. \eqref{HBLs}, of the averaged exchange energy $J\langle \boldsymbol{S}_i \rangle $ of conducting electrons on two different AFM sublattices. Since $J\langle \boldsymbol{S}_i \rangle \propto \boldsymbol{M}_i$, we get $E_{ex}=-\cal{J} (\boldsymbol{S_1}-\boldsymbol{S_2}) \propto -\cal{J} (\boldsymbol{M_1}-\boldsymbol{M_2}) = -\cal{J} \boldsymbol{L}$$_{AFM}$. This gives the left-side part of Eq. \eqref{EexB}. This analysis also shows that the low index $\sigma$ in the wave function $\psi_{\sigma}(r)$ is the projection of conducting-electron spin on the AFM order parameter $\boldsymbol{L}_{AFM}$.

Eqs. \eqref{dtau} and \eqref{EexB} lead to a  parabolic  isotropic negative magnetoresistance (NIMR)  persisting almost up to the 
  spin polarization field  $H_{\mathrm{sf}}$.   

{\large\bf Quantitative consideration}\\
Let us consider  quantitatively the electron WF squeezing in the AFM state.  The two AFM
sublattices are numerated by the index $j=1,2$,
and the spin projection on the AFM magnetization axis $\boldsymbol{L}_{AFM}$ is indexed by $\sigma =\pm 1\equiv \uparrow ,\downarrow $. For each quasi-momentum the quantum basis
consists of four states, $\left\vert j,\sigma \right\rangle  = \left\{ 1\uparrow ,1\downarrow ,2\uparrow
,2\downarrow \right\}$, corresponding to wave functions $\psi _{j,\sigma
}=\left\{\psi _{1\uparrow },\psi _{1\downarrow },\psi _{2\uparrow },\psi
_{2\downarrow }\right\} $. The usual Zeeman splitting $ E_{Z}(%
\boldsymbol{H})=\left( \boldsymbol{\vec{\sigma}\cdot \vec{H}}\right) g\mu
_{B}/2$  in a relevant external magnetic field $H\lesssim 5$T is much smaller
 than the exchange splitting  $ E_{ex0}\sim 30$meV and is
neglected below.  The neglected spin-orbit coupling $E_{SOC}\lesssim 5$meV (see Supplementary materials, Note 4) is also much smaller than the exchange splitting $E_{ex0}$ and, additionally, averages to zero after the integration over electron momentum in the first order of perturbation theory. Then for each electron quasi-momentum $\boldsymbol{k}$ the AFM-sublattice part of electron Hamiltonian
is given by the $4\times 4$
matrix, which decouples into two $2\times 2$ matrices: 
\begin{equation}
\hat{H}_{\sigma }=\left( 
\begin{array}{cc} 
 E_{ex}\sigma /2 & t_{0} \\ 
t_{0}^{\ast } & - E_{ex}\sigma /2%
\end{array}%
\right) .  \label{HBLs}
\end{equation}%
Here  the diagonal terms are the energy shifts due to the exchange interaction with AFM magnetization, and the  off-diagonal term $t_{0}=t_{0}^{\ast }$,  the intersublattice
electron transfer integral,  determines the electron hopping rate between AFM sublattices. The diagonalization of Hamiltonian (\ref{HBLs})
gives two eigenvalues $\varepsilon_{\pm ,\sigma }=\mp \sqrt{ E_{ex}^{2}/4+t_{0}^{2}}  $
and the corresponding normalized wave functions 
\begin{equation}
\psi _{\pm ,\sigma }=\frac{\psi _{1}\left( \sigma \gamma\pm \sqrt{\gamma
^{2}+1}\right) +\psi _{2}}{\sqrt{1+\left( \sigma \gamma\pm \sqrt{\gamma
^{2}+1}\right) ^{2}}},  \label{psipmWF}
\end{equation}%
where $\psi _{1},_{2}$ are the electron wave functions ``localized'' mostly on
the first and second AFM sublattices. 
Without the AFM order, i.e. at  $\gamma\propto  E_{ex}=0$, this gives the  simple electron
spectrum $\varepsilon_{\pm ,\sigma }^{0}=\mp t_{0}$ and
eigenstates $ \psi _{\pm ,\sigma }^{0}=\left( \psi _{2}\pm \psi _{1}\right) /\sqrt{2}=\psi_\pm^0$, 
which are the symmetric and antisymmetric superpositions of electron states on
sublattices $1$ and $2$ with equal WF weights on each AFM sublattice.
From Eq. (\ref{psipmWF}) we see that the AFM order lifts this symmetry,
 making the eigenfunction  amplitude larger 
 at one of the two sublattices at $\gamma \neq 0$
(see Fig. \ref{FigWF}). This enhances  the electron scattering rate in 
Eq.~(\ref{tau}), as seen from  Fig.~\ref{FigWF} and  estimated below. 

In the Born approximation, i.e. the second-order perturbation theory in
the impurity potential, the electron scattering rate is given by the Fermi's
golden rule \cite{Abrik}:%
\begin{equation}
\frac{1}{\tau _{\boldsymbol{n}} }=\frac{2\pi }{\hbar }\sum_{n^{\prime },i}\left\vert T_{ 
\boldsymbol{n}^{\prime }\boldsymbol{n}}^{(i)}\right\vert ^{2}\delta\left(
\varepsilon _{\boldsymbol{n}}-\varepsilon _{\boldsymbol{n}^{\prime }}\right)
,  \label{TauGFR}
\end{equation}%
where the index  $\boldsymbol{n}\equiv \left\{ \boldsymbol{k},\beta ,\zeta ,\sigma \right\} $
numerates quantum states, $\beta $
numerates electron bands at the Fermi level coming from different atomic orbitals in multiband metal, $\zeta =\pm $ denotes two eigenstates in Eq. (\ref{psipmWF}),   $i$ numerates impurities, 
$\varepsilon _{\boldsymbol{n}}$ is the electron energy in state $n$, and $\delta (x)$ is the Dirac delta-function. The additional index $\zeta =\pm $ doubles the number of quantum states per each quasimomentum $\boldsymbol{k}$, which compensates the folding in half of the first Brillouin zone by the commensurate AFM.
 
The short-range impurities or other crystal defects in
solids are usually approximated by the point-like potential 
$V_{i}\left( \boldsymbol{r}\right) =U\delta\left( \boldsymbol{r}-%
\boldsymbol{r}_{i}\right) $.
Here we omit the spin index $\sigma $ because it is conserved by the potential scattering.  The corresponding matrix element of electron scattering is
$T_{\boldsymbol{n}^{\prime }\boldsymbol{n}}^{(i)}=U 
\Psi^{\ast } _{\boldsymbol{n}^{\prime }}\left( \boldsymbol{r}_{i}
\right) \Psi _{\boldsymbol{n}}\left( \boldsymbol{r}_{i}
\right) $, where $\Psi _{\boldsymbol{n}}\left( \boldsymbol{r}\right) =\psi _{\beta\zeta \sigma }\left( \boldsymbol{r}\right) \exp \left( i\boldsymbol{	kr}\right)$ is 
the electron Bloch wave function with periodic $\psi _{\beta\zeta \sigma }\left( \boldsymbol{r}\right) $. 
For  point-like potential  the matrix element $T_{\boldsymbol{n}^{\prime }\boldsymbol{n}}^{(i)}$ does not depend on electron
momentum $\boldsymbol{k}^{\prime }$. Then the summation over $\boldsymbol{k}%
^{\prime }$ with the $\delta $-function in Eq. (\ref{TauGFR}) gives the
electron density of states (DoS) $\nu _{F\beta\zeta  \sigma}\equiv \nu _{\beta\zeta  \sigma}\left(
\varepsilon _{F}\right) $  at the Fermi energy $\varepsilon _{F}$ per one
spin component $ \sigma$ and per one subband. The total DoS per one
spin $\nu _{F}=\sum_{\beta\zeta }\nu _{F\beta\zeta \sigma}$ does not depend on $\sigma$ 
because the total time-reversal symmetry is conserved by AFM, as the simultaneous change of spin index $\sigma =\pm 1$
and of AFM sublattice $1 \leftrightarrow 2$ (layer in our case) does not change the system.
If the impurities  are
uniformly and randomly distributed in space, the sum over impurities
rewrites as an integral over impurity coordinate \cite{Abrik}: $\sum_{i}\rightarrow
n_{\mathrm{imp}}\int d^{3}\boldsymbol{r}_{i}$, where $n_{\mathrm{imp}}$ is the impurity
concentration. Then Eq. (\ref{TauGFR}) becomes 
\begin{equation}
\frac{1}{\tau _{\boldsymbol{n}} }=\frac{2\pi }{\hbar }n_{\mathrm{imp}}U^{2}\nu _{F}\,I,  \label{Tau2}
\end{equation}%
where the integral over one elementary cell
\begin{equation}
I\equiv \int d^{3}\boldsymbol{r}\left\vert \psi _{\beta\zeta  \sigma}\left( \boldsymbol{r%
}\right) \right\vert ^{2}\sum_{\beta^{\prime } \zeta ^{\prime } }\frac{\nu _{F\beta^{\prime } \zeta ^{\prime
} \sigma}}{\nu _{F}}\left\vert \psi _{\beta^{\prime } \zeta ^{\prime } \sigma}\left( \boldsymbol{r}\right)
\right\vert ^{2} . \label{I}
\end{equation}%

 According to Eq. (\ref{psipmWF}), for each band $\beta$ the WFs of two opposite subbands $\zeta =\pm$ shift toward opposite AFM sublattices. These states are separated by large energy
\begin{equation}
	\Delta \varepsilon_{\zeta }=\varepsilon_{-}-\varepsilon_{+}=2\sqrt{ E_{ex}^{2}/4+t_{0}^{2}}>2t_0.
	\label{DEpm}
\end{equation}%
 Hence, rather commonly a single subband $\zeta $ appears at the Fermi
level, i.e. $\nu _{F\beta\zeta ^{\prime }\sigma}=0$ for $\zeta ^{\prime }\neq
\zeta $. Then Eqs. (\ref{Tau2}),(\ref{I}) confirm Eq. (\ref{tau}). In fact, it holds even if there are several conducting bands $\beta$ but 
only one subband $\zeta$ is occupied, because the WF in Eq. (\ref{psipmWF}) does not considerably depend on the band $\beta$ and enters in the fourth power in Eq.~(\ref{I}). 
 Eqs.~(\ref{Tau2}),(\ref{I}) approximately
give Eq.~(\ref{tau}) even when there are several subbands $\zeta=\pm $ on the Fermi level, but their DoS $\nu _{F\beta \zeta \sigma}$ differ strongly. The latter happens in Dirac semimetals, where the
DoS $\nu _{\beta\zeta 
}\left( \varepsilon _{F}\right) \propto \varepsilon _{F}^{d-1}$ strongly
depends on energy $\varepsilon _{F}$ and where the subband splitting  $
\Delta \varepsilon_{\zeta } \gtrsim \varepsilon _{F}$. 

 For the reason explained above, we now consider the case of only one subband $\zeta=+$ at the Fermi level and estimate the difference $\delta I$ of two integrals (\ref{I}) with and without AFM, i.e.  for $\gamma \neq 0$ and $\gamma = 0$. 
The host crystal lattice has the $\hat{T}_2$ symmetry,  
therefore $\int \psi _{1}^{4}\left( z\right) dz=\int \psi _{2}^{4}\left( z\right) dz$.
For simplicity, we also assume that the overlap of the wave functions on
different AFM sublattices is negligible, i.e. $\psi _{1}\psi _{2}\ll
\left\vert \psi _{1}\right\vert ^{2}$, and we neglect the product $\psi
_{1}\psi _{2}\approx 0$. The main conclusion remains valid also when $\psi
_{1}\psi _{2}\sim \left\vert \psi _{1}\right\vert ^{2}$, but the calculations
are more cumbersome.

 Without AFM order, substituting  Eq. (\ref{psipmWF}) at $\gamma = 0$ to the integral (\ref{I}), we obtain $I_0=\int d^{3}\boldsymbol{r\,}\left\vert \psi
_{+}^{0}\right\vert ^{4}dz\approx \int d^{3}\boldsymbol{r\,}\psi
_{1}^{4}/2 $.
The difference $\delta I\equiv I-I_0=\int d^{3}\boldsymbol{r\,}\left( \left\vert
\psi _{+}\right\vert ^{4}-\left\vert \psi _{+}^{0}\right\vert
^{4}\right) $
determines the correction to electron mean free time $\tau $ according to Eqs. (\ref%
{tau}) or (\ref{Tau2}) and,   hence, the NIMR effect. After substituting Eq. (\ref{psipmWF})
at $\psi _{1}\psi _{2}\ll \left\vert \psi _{1}\right\vert
^{2}$ it reduces to
\begin{equation}
\delta I \approx \boldsymbol{\,}\frac{\gamma ^{2}}{ 1+\gamma ^{2}}
\int d^{3}\boldsymbol{r}\frac{\psi _{1}^{4}\left( z\right) }{2}=
\frac{\gamma ^{2}\,I_0}{ 1+\gamma ^{2}} =
\frac{\gamma ^{2}\,I}{ 1+2\gamma ^{2}}.  \label{dI}
\end{equation}%

At $\gamma ^{2}\ll 1$ Eq. (\ref{dI}) simplifies to $\delta I\approx \gamma ^{2}I_0\approx \gamma ^{2}I$,
and substituting  Eqs. (\ref{dI}) and (\ref{EexB}) to (\ref{Tau2}) we obtain the relative
increase of resistivity due to the AFM ordering%
\begin{equation}
\frac{\delta \rho (H)}{\rho (0)}
= \frac{\delta I}{I}= \frac{\gamma^{2}}{1+2\gamma^{2}} \approx \left( \frac{
E_{ex0}}{2t_{0}}\right)^{2}\left( 1-\frac{H^{2}}{H_{\mathrm{sf}}^{2}}%
\right).  
\label{drho}
\end{equation}%
 Contrary to the NMR caused by   chiral anomaly in Weyl semimetals and to the GMR in layered heterostructures, this increase of
resistivity is {\em isotropic}. 
Therefore, our NIMR mechanism applies both for the interlayer and intralayer transport in layered conductors,
and only slightly depends on the magnetic field direction due  to a 
magnetic anisotropy solely.\\

{\large\bf Classical positive magnetoresistance}\\
At $H>H_{\mathrm{sf}}$ the obtained NIMR correction (\ref{drho}) disappears, and
for the current $\boldsymbol{J}$ perpendicular to magnetic field direction one returns to the usual classical 
positive magnetoresistance (CPMR)  in multiband conductors due to
impurity scattering, which is parabolic at low field when $\omega_c \tau \ll 1 $ \cite{PRB_tbp}: 
\begin{equation}
\rho_{zz}^{m}\left( H\right) /\rho_{zz}^{m}\left( 0\right) \approx 1+\omega_c
^{2}\tau ^{2},\ \ \omega_c \tau \ll 1,  
\label{RMzB}
\end{equation}
where $\omega_c =eH/(m^{\ast}c)$ is the cyclotron frequency and $\tau$ - the scattering time for the dominant band  \cite{PRB_tbp}. 

\begin{figure}[tbh]
\includegraphics[width=0.5\textwidth]{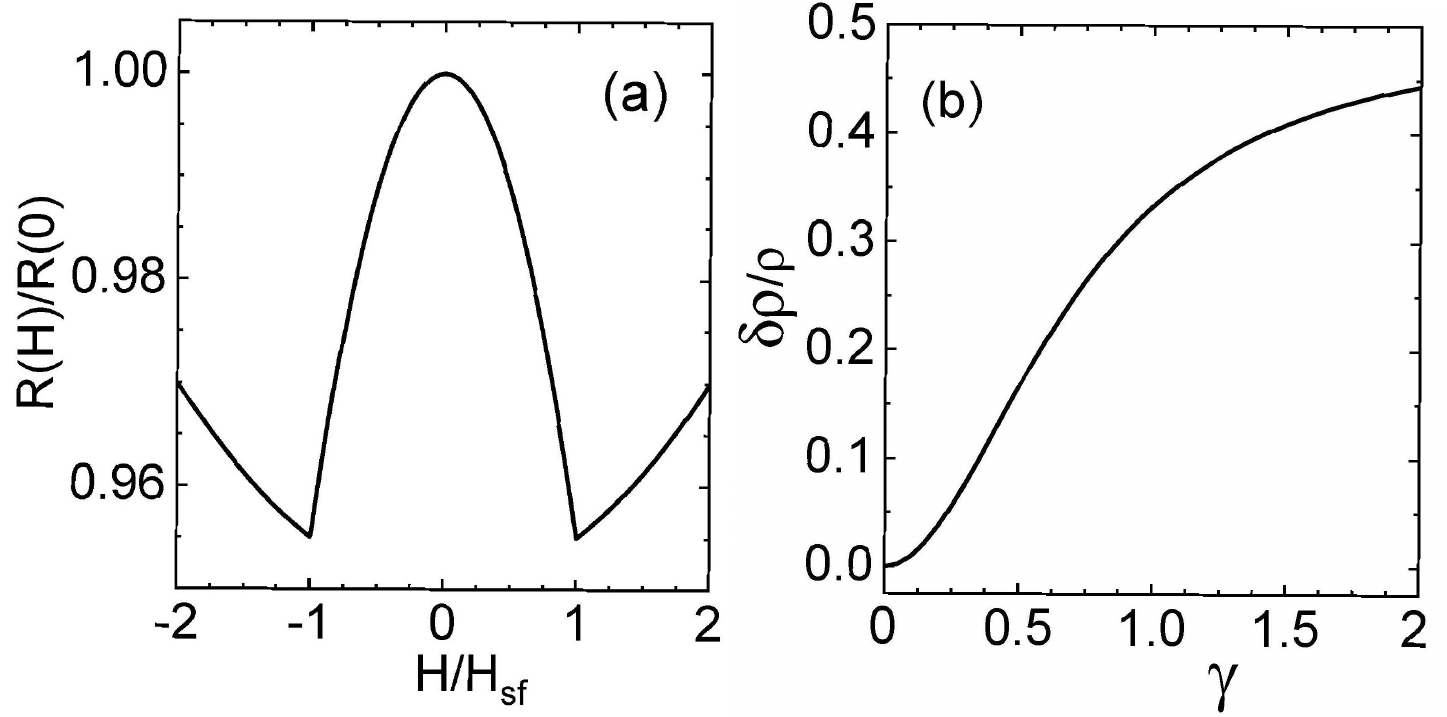}
\caption{{\bf Calculated negative magnetoresistance.} 	{\bf a} Magnetoresistance curve given by   Eqs. (\protect\ref{drho}) and (\ref{RMzB})  at $%
 E_{ex0}/2t_z =0.23$ and $[\protect\omega_c\protect\tau 
(H=H_{\mathrm{sf}})]^2 =0.2$. 	{\bf  b} The maximal possible relative value 
of the proposed NIMR effect as a  function of $\gamma_0= E_{ex0}/2t_{0} $, plotted using Eqs.  (\ref{dI}) or (\ref{drho}). 
It saturates at $50$\%  for $\gamma_0\gg 1$, when resistivity drops by half.
}
\label{FigRB}  
\end{figure}

\begin{figure}[ht]
	\includegraphics[width=0.4\textwidth]{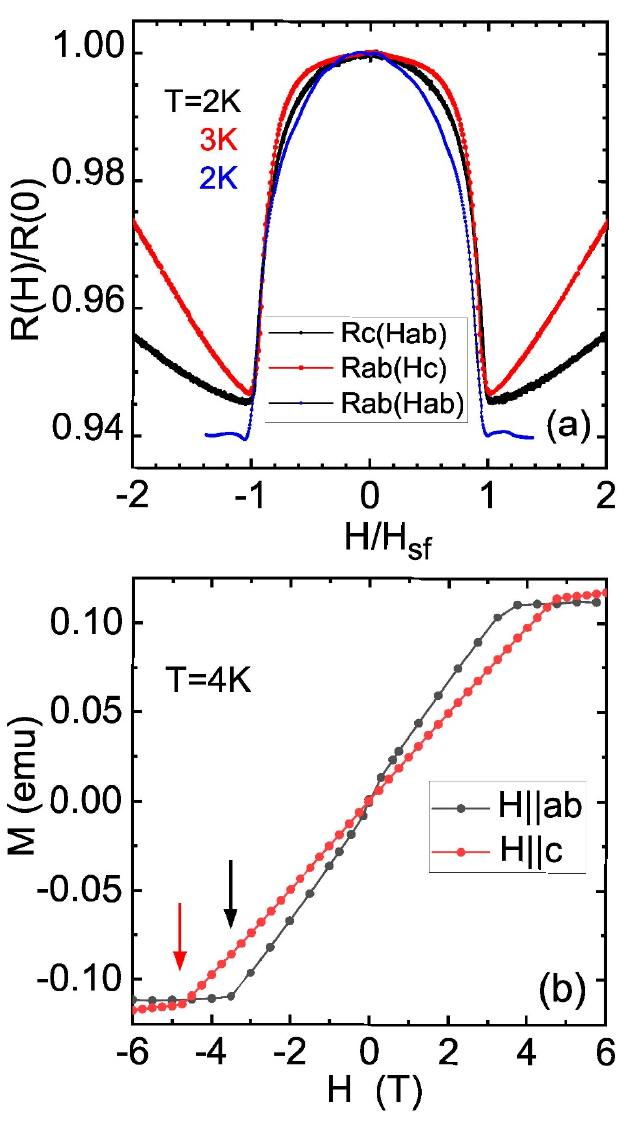}
	\caption{{\bf Experimental data on magnetoresistance and magnetization of EuSn$_2$As$_2$ bulk crystal.} {\bf a} Examples of the normalized magnetoresistance  $R(H)/R(0))$ vs normalized field $(H/H_{\rm sf})$ for two orientations of the bias current, in-plane $R_{ab}$ and perpendicular to the plane $R_c$, 
		and for two magnetic field directions, $\boldsymbol{H}\|(ab)$ and $\boldsymbol{H}\|c$.  Data are taken at $T=2-3$K. {\bf b} Magnetization $M(H)$  dependences  for two field orientations, $\boldsymbol{H}\|ab$ and $\boldsymbol{H}\|c$. The nonlinearity of $M(\boldsymbol{H}\|ab)$ in low fields is related with   spin canting and spin-flop \cite{PRB_tbp}. Vertical arrows depict $H_{\rm sf}$ value for two field orientations.
	}
	\label{fig:parabolic_MR}
\end{figure}

Combining Eqs. (\ref{drho}) and (\ref{RMzB}) gives the schematic MR curve illustrated in Fig.~\ref{FigRB}a, which resembles very much  typical experimental   data 
 measured with the representative compound EuSn$_2$As$_2$ (see Fig.~\ref{fig:parabolic_MR}a).  The sharp transition from negative to positive magnetoresistance at $H=H_{\mathrm{sf}}$ happens because our NIMR effect, which is stronger than the classical positive magnetoresistance in A-type AFM layered metals, exists only in the AFM state at $H<H_{\mathrm{sf}}$. 
 
 It is worth of noting, the simplified model, Eq.~(\ref{drho}), is the lowest order theory. 
 Closer to the 	spin polarization 
 transition, at $H\rightarrow H_{\mathrm{sf}}$,  higher-order magnetization fluctuation effects may cause a faster decrease of $L_{AFM}\left( H\right) $ than that given by Eq. (\ref{EexB}).

{\large\bf Comparison of the model with experimental data}\\
In isotropic 3D AFM metals, the proposed NIMR mechanism is usually very weak, because the
ratio $\gamma = E_{ex}/2t_{0}\ll 1$. Indeed, $ E_{ex}\lesssim
0.1$eV, while $2t_{0}\sim 1$eV is comparable to the bandwidth. However, in
strongly anisotropic layered AFM  semimetals 
the ratio $\gamma = E_{ex}/2t_{0}\sim 1$, because the
interlayer transfer integral $t_{0}\lesssim 0.1$eV in van-der-Waals or other layered compounds is also small.

To compare the presented theory with experimental data,  we performed magnetization and magnetoresistance 	measurements with EuSn$_2$As$_2$ bulk single crystals. 
  
The normalized $R(H)/R(0)$ magnetoresistance and magnetization $M(H)$ dependences 
 are shown in Figs.~\ref{fig:parabolic_MR}  for various field and current directions, $H\|c$,  $H\|(ab)$, $J\|c$, $J\|(ab)$,
 at our lowest temperature $T\approx 2\,{\textrm K} \ll T_N\approx 24$\,K. More detailed magnetization data for various temperatures are presented in Fig.~S1 of Supplementary Note~1.
 
As field exceeds $H_{\mathrm{\rm sf}}$, the NIMR hump changes sharply to the  conventional smooth  parabolic rise. 
The sharp transition between the negative 
and positive 
magnetoresistance (i.e. the $R(H)$ minimum) coincides with the sharp magnetization saturation at $H=H_{\mathrm{sf}}$ for both field
directions (Figs.~\ref{fig:parabolic_MR}a,b),
consistent also with previous studies \cite{chen_ChPhysLet_2020, li_PRB_2021, sun_SciChin_2021, pakhira_PRB_2021, golov_JMMM_2022}.
Comparing this data with the presented first-order in $(\omega_c\tau)^2$ model, 
we conclude that the model correctly captures the main features: approximately parabolic NMR, 
its magnitude, and  a sharp transition to the conventional CPMR. One can also see in Fig.~\ref{fig:parabolic_MR}a that 
CPMR is much weaker for
$\boldsymbol{J}$  perpendicular to the layers, than along the $ab$ plane, and is almost missing for $\boldsymbol{J\| \boldsymbol{H}}$. This observation is in a qualitative agreement with the simple treatment of CPMR,  Eq.~(\ref{RMzB}), and with zero-field conduction anisotropy in van der Waals (vdW)-type layered crystal \cite{PRB_tbp}. The observed NIMR is larger for $\boldsymbol{J\| \boldsymbol{H}}$ than for $\boldsymbol{J\perp \boldsymbol{H}}$ just by the magnitude of this CPMR.

The measured NIMR (see Fig.~\ref{fig:parabolic_MR}a and also refs.~\cite{gui_ACSCentSci_2019, ying_PRB_2012, yang_JMMM_2019}) looks somewhat more flat than the parabolic dependence (Eq.~\ref{drho} and Fig.~\ref{FigRB}a). We presume that  flattening of the data  in low fields may be caused by magnon scattering that in AFM state results in a positive MR \cite{yamada-takada_1973-1, yamada-takada_1973-2}. This scattering mechanism is  beyond the framework of  our  first order theory. 

 In our measurements   on EuSn$_{2}$As$_{2}$ crystals, 
 the magnetoresistance   $\delta \rho /\rho$ drops by about $5-6\%$, as shown in Fig. \ref{fig:parabolic_MR}. According to Eq. (\ref{drho}) 
  this  drop  corresponds to 
\begin{equation}
 \gamma_0 = E_{ex0}/2t_{0}=\sqrt{\delta \rho /\rho }\approx 0.23.  \label{ratio}
\end{equation}%
We now compare this ratio with our DFT calculations and ARPES data for $ E_{ex0}$ and 
$4t_{0}$. 
From the DFT calculations we found  $ E_{ex0}\approx  30$\,meV level splitting
for EuSn$_{2}$As$_{2}$ \cite{PRB_tbp}. Substituting this to Eq. (\ref%
{ratio}) one gets $t_{0}\approx 65$meV. The 
ARPES data \cite{li_PRX_2019, PRB_tbp} do not have sufficient energy resolution to measure the bilayer 
splitting and $t_0$ directly. However, the $t_0$ value can also be 
roughly estimated from the observed resistivity anisotropy  $\rho _{zz}/\rho
_{xx}\approx 130$  \cite{PRB_tbp} and the width of the in-plane energy band $4t_{x}\approx 1.9$%
eV, which is taken from the ARPES data and from the DFT calculation. As a
result, we obtain the lowest estimate  $t_{0}\approx t_{x}/\sqrt{\rho
_{xx}/\rho _{zz}}\approx 42$meV, which is in a reasonable agreement with $%
t_{0}\approx 65$meV determined from Eq. (\ref{ratio}).

\begin{figure}[tbh]
	\includegraphics[width=0.48\textwidth]{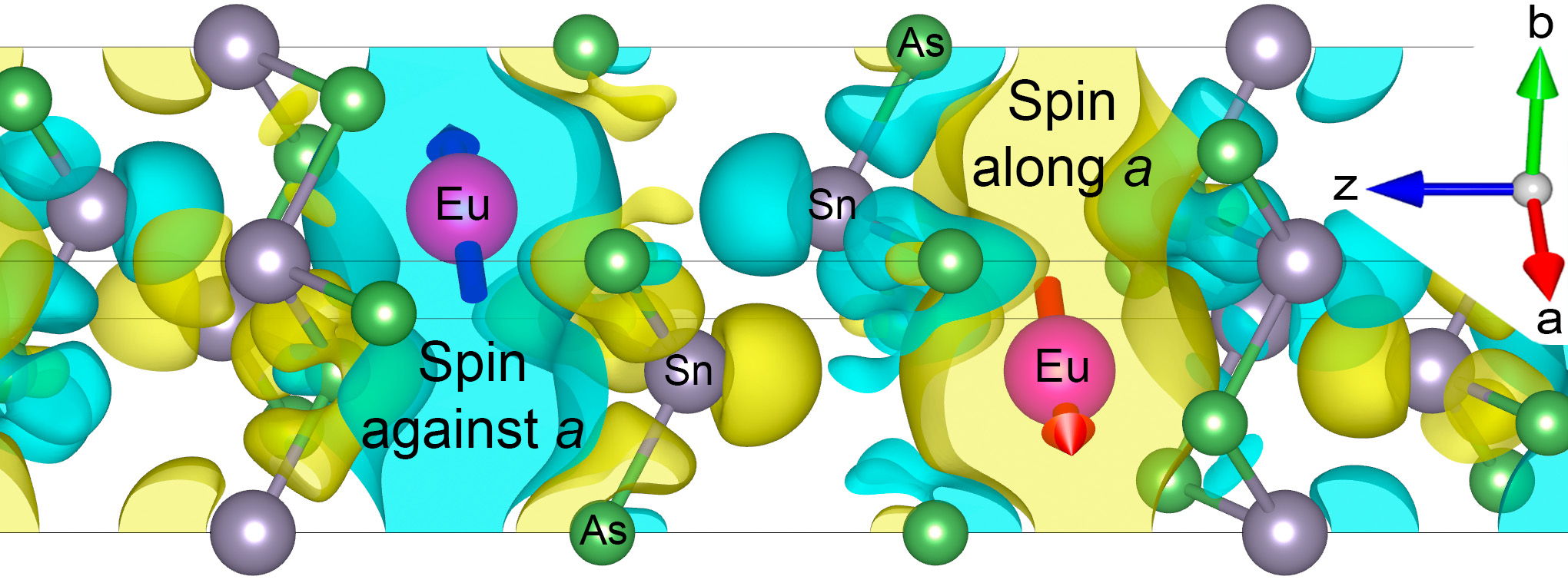}
	\caption{{\bf Calculated spin-polarized electron density distribution.} 
			The yellow (turquoise) isosurfaces correspond to fixed differences of charge density with spin along (against) the $a$ axis of AFM ordering of Eu spins. The neighboring SnAs layers have opposite spin polarization of electrons at the Fermi level thus substantiating our model.}
	\label{FigDFT}
\end{figure}

\vspace{0.1in}{\Large\bf Discussion}\\
Now we check whether our assumption of complete $\zeta$-subband polarization is valid for
EuSn$_{2}$As$_{2}$. 
According to the DFT calculations  \cite{PRB_tbp}, in EuSn$_{2}$As$_{2}$ there are two
electron bands with $\varepsilon _{F}\approx 55$ meV and $\varepsilon
_{F}\approx 100$ meV and two hole bands with $\varepsilon _{F}\approx 65$
meV and $\varepsilon _{F}\approx 80$ meV. 
Substituting $t_{0}\approx  50$meV and $%
 E_{ex0}\approx 30$meV to Eq. (\ref{DEpm}) we get the $\zeta$-subband splitting $%
\Delta \varepsilon_{\zeta }\gtrsim 100$meV$>\varepsilon _{F}$ for all Fermi-surface
pockets.  Hence, in EuSn$_{2}$As$_{2}$ all electronic bands at the Fermi level are completely
subband-polarized, and the above analysis  of the NIMR effect 
is applicable to the experimental data in EuSn$_{2}$As$_{2}$.  

In order to substantiate our model of opposite spin polarization of conducting electrons on neighboring SnAs layers, we performed 
the ab-initio DFT calculation of spin-polarized electron density distribution.
Our results are shown in Fig. \ref{FigDFT}, where the yellow (turquoise) isosurfaces limit the real-space areas filled 
with the same color of positive (negative) electron spin density exceeding a fixed absolute value. As we assumed in our 
model and illustrated in Fig. \ref{FigWF}, the neighboring SnAs layers indeed have opposite spin polarization of conducting electrons. 
This spin polarization in EuSn$_2$As$_2$ is small, $\sim 0.1$\% of total electron density, but it is peaked at the Fermi level and strongly affects the electron scattering rate according to the above analysis.

The spin-orbit coupling (SOC) is ignored in our model because the corresponding term $ E_{SOC}$ in electron energy is usually much smaller than the exchange coupling $ E_{ex0}$. A weak spin-orbit coupling gives a perturbative correction to the predicted NIMR effect, which is small by the parameter $ E_{SOC}/ E_{ex0}\ll 1$ and does not affect our NIMR mechanism considerably. Moreover, the spin-orbit term in the Hamiltonian is averaged over electron momenta to almost zero in the first order of perturbation theory and affects only in its next orders. Hence, even at $ E_{SOC}/ E_{ex0}\sim 1$ our NIMR mechanism survives, because it gives the effect averaged over all electron quasi-momenta and contains only the exchange splitting. Of course, at $ E_{SOC}/ E_{ex0}\gtrsim 1$ our NIMR mechanism quantitatively modifies, which requires further investigation. However, in the studied compound EuSn$_2$As$_2$ the exchange splitting $ E_{ex0}\approx 30$meV, while the spin-orbit coupling $E_{SOC}\lesssim 5$\,meV$\ll  E_{ex0}$, as follows from our DFT calculations (see {\bf Supplementary Note 4}). Hence, we can safely neglect the SOC. 

The influence of the redistribution of spin-density of conduction electrons on the exchange interaction between the Eu atoms is also ignored. This redistribution may only slightly affect the RKKY interaction between the Eu atoms. The redistribution of spin density of conduction electrons by AFM order is small by parameter $\gamma \ll 1$, while their charge density remains practically unchanged. Moreover, the main contribution to the exchange interaction between the Eu atoms comes from the direct exchange and superexchange coupling via intermediate localized electrons of Sn and As orbitals rather than from the RKKY mechanism. This can be estimated directly or concluded from the weak dependence of the AFM coupling on the electron density tuned by pressure \cite{sun_SciChin_2021} in EuSn$_2$As$_2$ or by doping in a sister compound EuIn$_2$As$_2$ \cite{EuIn2As2Jian_2024}.

Our  ARPES measurements and band structure calculations for bulk EuSn$_2$As$_2$ \cite{PRB_tbp} 
	show no band crossings and Dirac points at the Fermi level (see Supplementary Note 4), and are consistent with calculations of 
	Ref.~\cite{arguilla_InChemFront_2017}. The magnetotransport measurements also showed that   $\delta R(H)/R(0)$ is almost independent 
 of the field direction $H\|ab$ or $H\|c$, of the charge transport direction,
in-plane $R_{ab}(H)$ or normal to the plane $R_c(H)$ (see Fig.~\ref%
{fig:parabolic_MR} and Ref.~\cite{PRB_tbp}).  NIMR is also independent of the angle between the 
electric and magnetic field in the easy $ab$-plane \cite{PRB_tbp} (see {\bf Supplementary Note 5}).  These results exclude the 
 axion insulator origin of NIMR in EuSn$_2$As$_2$. 
 The similarity of NIMR in various materials also  indicates insignificance of the 
 Dirac states close to $E_F$, such as in  EuSn$_2$P$_2$  \cite{gui_ACSCentSci_2019}. 
 
Neither our, nor other ARPES measurements and DFT band structure calculations for EuSn$_2$As$_2$  \cite{PRB_tbp, arguilla_InChemFront_2017, lv_ASCAplElextronMat_2022} reveal Dirac cones at $E_F$.
In addition, resistivity temperature dependence for EuSn$_2$As$_2$ in all works 
\cite{pakhira_PRB_2021, chen_ChPhysLet_2020,li_PRB_2021, PRB_tbp} shows a monotonic decrease almost down to $T_N$, 
thus evidencing a conventional semimetalic-, rather than topological insulator-type behavior, such as suggested in Ref.~\cite{li_PRX_2019}.
These results provide solid ground for the applicability of our model to  EuSn$_2$As$_2$.

The  independence of NIMR on the sample thickness 
(from 60\,nm to 0.2\,mm) \cite{maltsev_tbp} 
indicates that NIMR does not come from scattering by large-scale
lattice defects (such as misfit dislocations, misoriented grains), 
and is irrelevant to the zero-field electron scattering rate $1/\tau_0$, since these parameters are 
to be different for various samples, for the bulk crystal and  exfoliated flakes a few monolayer thick. 
Whereas the NIMR   magnitude is about  the same for all studied  samples and for various field directions, its shape 
for some samples in low fields  looks somewhat more flattened, 
that might be caused by  a minor positive MR contribution due to magnon scattering \cite{PRB_tbp}


\begin{figure*}
	\includegraphics[width=\textwidth]{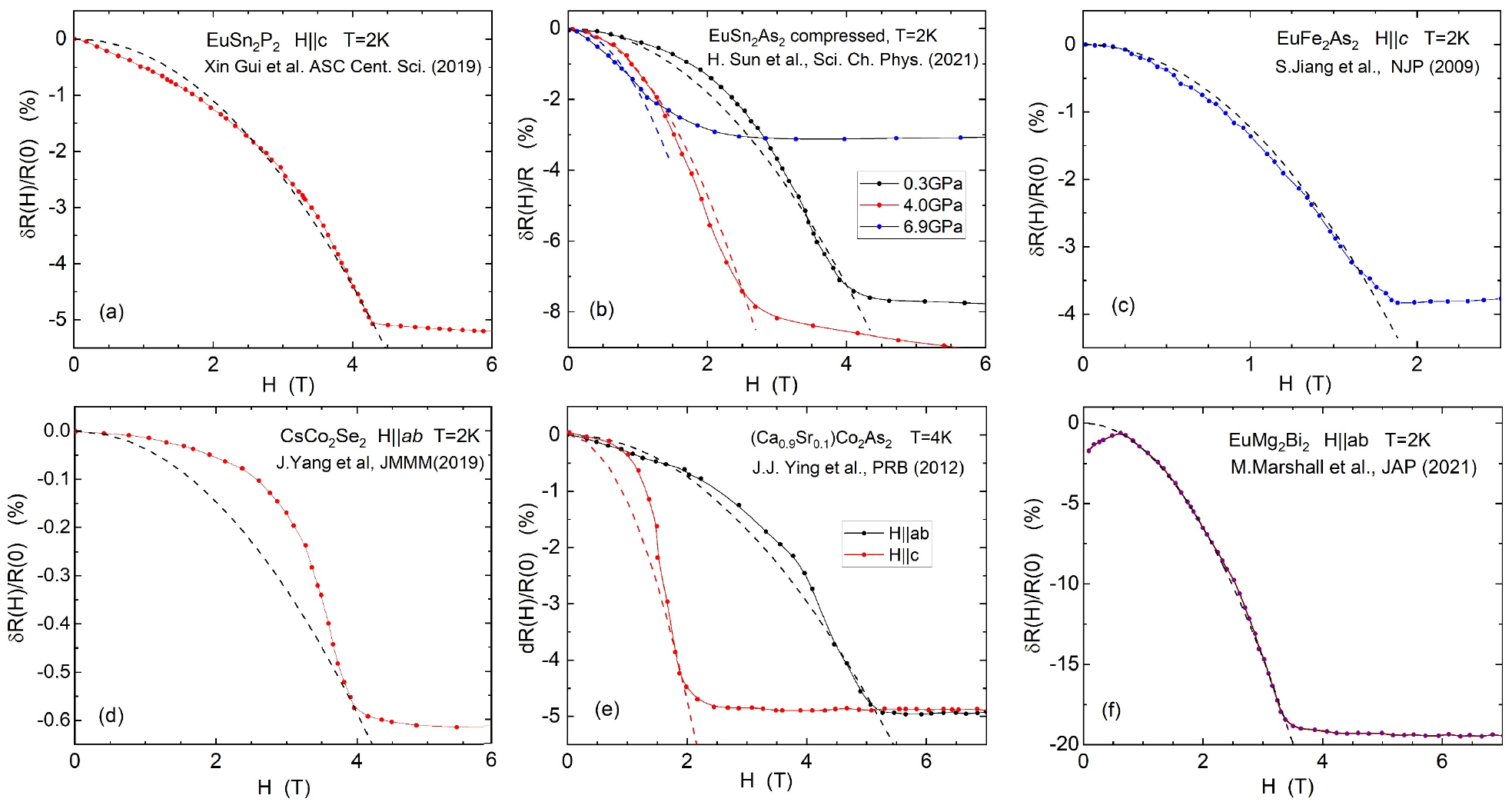}
	\caption{ {\bf Magnetoresistance $\delta R(H)/R(0)$ for several layered vdW materials in the AFM state.} The experimental data are digitized from the published papers (symbols), and model curve are plotted according to Eq.\,(\ref{drho}) from the main text (dashed lines) for several compounds: (a)  for EuSn$_2$P$_2$ from 	Fig.\,5d of \cite{gui_ACSCentSci_2019}, (b) for the compressed EuSn$_2$As$_2$ from Ref.~\cite{sun_SciChin_2021} at several pressure values indicated on the panel, (c) for EuFe$_2$As$_2$ from Fig.\,6 of Ref.~\cite{jiang_NJP_2009}, (d) for CsCo$_2$Se$_2$ from Fig.\,4 of Ref.~\cite{yang_JMMM_2019}, (e) for (CaSr)Co$_2$As$_2$ from Fig,\,4 of Ref.~\cite{ying_PRB_2012}, (f) for EuMg$_2$Bi$_2$ from Fig.\,3 of Ref.~\cite{marshall_JAP_2021}. }
	\label{analogues}
\end{figure*}
	
{\bf The proposed magnetoresistance model is widely universal.} 
 The approximately parabolic and isotropic negative magnetoresistance  measured in this work is
similar to that observed earlier in EuSn$_2$As$_2$ \cite{sun_SciChin_2021, li_PRB_2021}, 
and in sister materials -- EuSn$_2$P$_2$ \cite{gui_ACSCentSci_2019}, CaCo$_2$As$_2$ \cite{ying_PRB_2012}, and CsCo$_2$Se$_2$ \cite{yang_JMMM_2019}.  
We, therefore, believe that our model is applicable to the listed layered AFM-ordered  compounds.
It  may also be partially applicable to EuFe$_2$As$_2$, though 	the in-plane properties of  this compound  are
	affected by Fe-moments stripe ordering \cite{jiang_NJP_2009, sanchez_PRB_2021}, which causes  anisotropy of  NMR.

 In order to illustrate  the universality of the proposed mechanism, we show below 
 the magnetoresistance $\delta R(H)$ data for several vdW layered compounds in the AFM state (i.e. at $T<T_N$) and compare them qualitatively 
 with Eq.~(\ref{drho}). In all cases, the MR data scale with field $H_{\rm sf}$ of magnetization saturation, where the $H_{\rm sf}$ value is taken from the magnetization measurements. As for the magnitude of NIMR effect, since  $( E_{ex0}/2t_0)$-value is unavailable  in the published papers, it was used as a adjusting parameter to fit $\Delta R(H)/R(0)$ at the point $H=H_{\rm sf}$. In all compounds this parameter takes reasonable values. One can see that the predicted parabolic dependence $\delta R(H)$ qualitatively fits the data; the deviation (e.g., on  panel d) may be associated with oversimplified character of the lowest-order  model that ignores, e.g.,  the positive MR from electron-magnon scattering (for discussion, see Ref.~\cite{PRB_tbp} and the references therein).

 It is worth noting,  the panel  (f) for  EuMg$_2$Bi$_2$  shows a more complex non-monotonic  MR in a weak field $H\ll H_{\rm sf}$ in addition to the predicted NIMR effect. Nevertheless, the abrupt termination of NIMR at $H=H_{\rm sf}$ is firmly connected with the magnetization saturation, as well as in all listed above AFM layered semimetals, in agreement with our model. Concerning the low-field $R(H)$ feature in EuMg$_2$Bi$_2$, we note firstly that the magnetoresistance rise in low field is observed for the in-plane transport with magnetic field also aligned in the basal plane. Secondly, the positive MR  slowly broadens but does not decay with temperature increasing up to about \cite{marshall_JAP_2021} 100\,K and, finally, it overpowers NIMR at $T>20$\,K. These two facts do not allow to associate the positive MR at $H\ll H_{\rm sf}$ with  antilocalization. 
 The  EuMg$_2$Bi$_2$ material is considered to be the nodal-line semimetal with linear-dispersing bands at the Fermi level \cite{marshall_JAP_2021}. Since in this AFM-material the FM-type spin fluctuations have been found to dominate, the sign-changing MR was discussed in Ref. \cite{marshall_JAP_2021} in terms of  the  competition between the AFM and FM orders.
 Also, for EuMg$_2$Bi$_2$ the spin-orbit coupling is found to be crucially essential for the band structure at the Fermi level and makes the topological invariant equal to 1; therefore, the influence of non-trivial topology on magnetotransport should be important. Obviously, complex magnetoresistance behavior in this interesting material requires detailed studies.

 There are other even more complex behaviors of magnetoresistance in more complex layered materials, for example, in  
 ferromagnetic Weyl semimetal  EuCd$_2$As$_2$ \cite{roychowdhury_AdvSci_2020}, in axion insulator EuIn$_2$As$_2$ 
 \cite{yu_PRB_2020, yan_PRR_2022}, and AFM topological insulators EuMg$_2$Bi$_2$ \cite{marshall_JAP_2021},  MnBi$_2$Te$_4$  and MnBi$_4$Te$_7$ \cite{ge_NatSciRev_2020, tan_PRL_2020}. 
 These cases are beyond the framework of suggested simple model,  but even in these compounds our mechanism may considerably contribute to the observed negative magnetoresistance, e.g., as we see from Fig. \ref{analogues}f.\\

{\bf To conclude,} we clarified the origin of the \emph{negative isotropic}
magnetoresistance observed in layered AFM semimetals. Specifically,  we 
suggested a novel type of  magnetoresistance mechanism and
developed a theory describing NIMR over a wide field
range up to complete spin polarization. In the
proposed theory, the negative magnetoresistance originates from the 
violation of $\hat{T}_2$ symmetry and the corresponding stronger localization of 
electron wave functions on one of the two AFM sublattices depending on electron spin.

The proposed mechanism of the parabolic NIMR  is generic for the layered AFM  semimetals;  
 the NIMR magnitude is independent of the electron scattering rate and increases 
as the ratio of the exchange splitting to the transfer integral increases. 
The proposed model agrees qualitatively with the available data for layered AFM semimetals, such as 
EuSn$_2$As$_2$, EuSn$_2$P$_2$, EuFe$_2$As$_2$, CaCo$_2$As$_2$, CsCo$_2$Se$_2$,
confirming negative isotropic magnetoresistance to be their intrinsic property, irrelevant to defects, 
domains, and other sample-specific disorder. Although  the NIMR magnitude is about the same in all of the above compounds,
$\approx 5-6\%$, their relevant  energy spectrum parameters ($ E_{ex0}$ and $t_0$)   are unknown, which prevents a detailed comparison with theory.

 The proposed model of the NIMR effect helps to extract useful information about the electronic structure of AFM compounds.
Indeed, according to Eq. (\ref{drho}),
the NIMR magnitude gives the parameter $\gamma_0 = E_{ex0}/2t_{0}$, i.e. the ratio of exchange energy splitting to the hopping amplitude between AFM sublattices. 
The proposed NIMR effect opens a platform for the detection and study of magnetic ordering using electron transport measurements.

\vspace{0.1in}
{\Large\bf Methods}\\
{\bf Experimental}. The magnetotransport measurements 
were preformed with CFMS-16 cryomagnetic system,
and using standard 4-terminal AC technique with SR830 lock-in amplifier.
Magnetization measurements 
were done with PPMS-7 SQUID magnetometer.

The  EuSn$_2$As$_2$ single crystals were synthesized from homogeneous
SnAs (99.99\% Sn + 99.9999\% As) precursor and elemental Eu (99.95\%)
in stoichiometric molar ratio (2:1) using the growth method, similar to
our previous works 
\cite{golov_JMMM_2022, eltsev_UFN_2014}. For  magnetization measurements
we selected bulk crystals of EuSn$_2$As$_2$,  $\approx 1-2$\,mm in the $ab$-plane and $\approx 0.1$\,mm thick; transport  
measurements were done both with same bulk crystals and with exfoliated flakes. 

\vspace{0.1in}{\bf DFT calculations.}
The DFT band structure calculations were performed within the DFT+U approximation in the VASP software package~\cite{vasp}. The generalized gradient approximation (GGA) in the form of the Perdew-Burke-Ernzerhof (PBE) exchange-correlation functional~\cite{DFT_PBE} was employed. The onsite Coulomb interaction of Eu-$4f$ electrons was described with the DFT+U scheme with the Dudarev approach~\cite{DFT_U} (U=5.0\,eV same as in Ref.~\cite{li_PRX_2019}).	

\vspace{0.1in}{\Large\bf Data availability}
The data that support the findings of this study are available from the
corresponding author upon reasonable request. 


\bibliography{PapersNIMR}

\section*{Acknowledgements}
The authors thank V.N. Menshov for useful discussions.
AVS, OAS,  and VMP were supported within  State Assignment of the research at LPI.
PDG acknowledges State Assignment \# FFWR-2024-0015. 
NSP, IAN and IRS acknowledge  partial support of the State Assignment
\# 124022200005-2 of Institute of Electrophysics and \# 124020600024-5 of Institute of Solid State Chemistry 
UB RAS. AVS, OAS,  VMP, NSP and IAN  acknowledge support 
from RSCF (grant \# 23-12-00307). Experimental work was partly done using equipment of LPI Shared facility Center.

\section*{Author contributions}
P.D.G. composed the theoretical model and performed analytical calculations,
N.S.P., I.A.N., and I.R.S. performed numerical DFT calculations.  
K.S.P., and E.M.,   grew and characterized the bulk crystals, 
V.M.P., A.V.S., and O.A.S. planned and designed the experiments, analyzed and processed the data.
A.V.S., and O.A.S.  performed magnetization and magnetotransport measurements with bulk samples.
E.M. prepared the flake samples, E.M. performed measurements with flakes and  processed their results.
P.D.G. and V.M.P. wrote the paper with considerable help from all authors. 
All the authors contributed to the discussion of the experimental results.


\section*{Additional Information}
The paper contains supplementary material. 
\medskip

{\bf Correspondence} and requests for materials should be addressed to
Pavel D. Grigoriev at grigorev@itp.ac.ru.
\end{document}


\title{{\it Supplementary Information for}\\ Universal negative magnetoresistance in antiferromagnetic metals caused by symmetry breaking of electron wave functions}

\author{Pavel D. Grigoriev$^*$}
\affiliation{L.D. Landau Institute of Theoretical Physics, RAS}
\affiliation{National University of Science and Technology ''MISiS'', 119049, Moscow, Russia} 
\affiliation{HSE University, Moscow 101000, Russia}
\author{Nikita S. Pavlov}
\affiliation{Institute for Electrophysics,  RAS, Ekaterinburg, 620016, Russia}
\affiliation{V.~L.~Ginzburg Research Center at P.~N.~Lebedev Physical Institute, RAS,  Moscow 119991, Russia}
\author{Igor A. Nekrasov}
\affiliation{Institute for Electrophysics, RAS, Ekaterinburg, 620016, Russia}
\author{Ilya R. Shein}
\affiliation{Institute of Solid State Chemistry, RAS, Ekaterinburg, 620990, Russia}
\author{Andray V. Sadakov}
\author{Oleg A. Sobolevskiy}
\affiliation{V.~L.~Ginzburg Research Center at P.~N.~Lebedev Physical Institute, RAS, Moscow 119991, Russia}
\author{Evgeny  Maltsev}
\affiliation{Leibniz Institute for Solid State and Materials Research, IFW Dresden, 
	D-01069 Dresden, Germany}
\affiliation{Dresden-W\"{u}rzburg Cluster of Excellence ct.qmat, Dresden, Germany}
\author{Vladimir M. Pudalov}
\affiliation{V.~L.~Ginzburg Research Center at P.~N.~Lebedev Physical Institute, RAS,  Moscow 119991, Russia}
\maketitle

\vspace{-24pt}
\begin{center}
$^*$Correspondence to: grigorev@itp.ac.ru
\end{center}
\vspace{12pt}
\tableofcontents
\newpage
\section{Supplementary Note 1:  Magnetization field dependence}
Our DC-magnetization measurements have been performed earlier with single crystal samples from the same batch \cite{golov_JMMM_2022}. 
Figure {\bf S1} shows results of magnetization measurements with EuSn$_2$As$_2$ (ESA) sample \cite{golov_JMMM_2022} in
two orientations of the sample in magnetic field. 
For the proposed  magnetoresistance model, it is essential only that as field increases, $M(H)$ curves  (at the lowest temperatures)  demonstrate a sharp saturation of the magnetization. The saturation field values are taken as complete spin  polarization field $H_{\rm sf}$. For different field orientations these values are different: $H_{\rm sf}= 3.6$\,T when magnetic field is applied in the $ab$ planes and $H_{\rm sf} = 4.9$\,T when magnetic field is applied along the $c$-
axis. In the field range $H\leq H_{\rm sf}$, magnetization $M$ develops linearly with field, thus substantiating  
our derivation  of  Eq.~(10) of the main paper.
The $M(T)$ curves (Figure S1c) \cite{golov_JMMM_2022} show the magnetic transition at temperature
$T_N \approx 24$\,K, manifested by the kink, which corresponds to the antiferromagnetic  spin ordering in the Eu sub-lattice
of ESA.
Both, the saturation fields and the critical temperature, 
 are well consistent with values reported for ESA
compound previously \cite{li_PRX_2019, chen_ChPhysLet_2020, pakhira_PRB_2021}.

\begin{figure*}[!ht]
	\begin{center}
	\includegraphics[width=0.47\columnwidth]{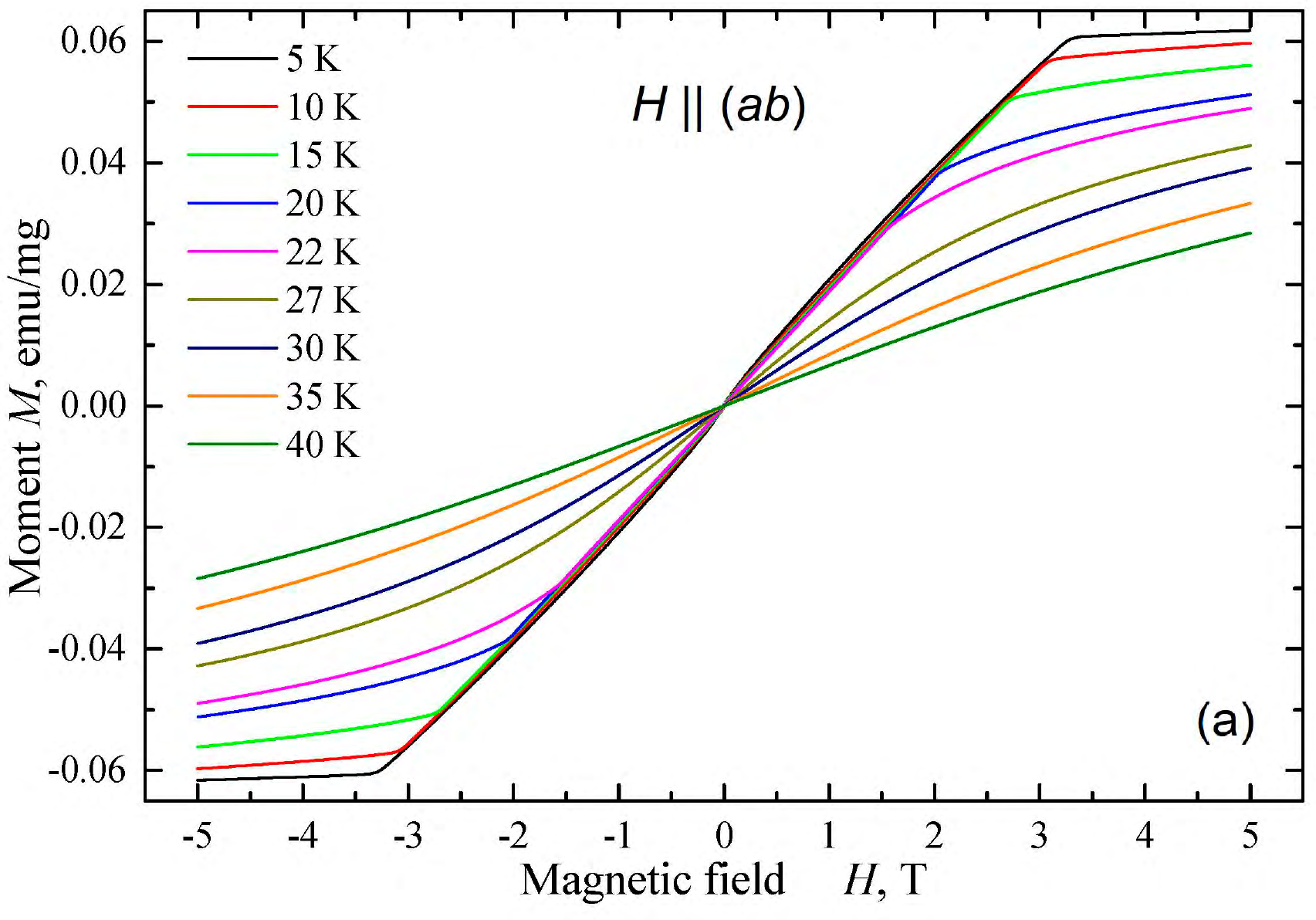}
	\includegraphics[width=0.47\columnwidth]{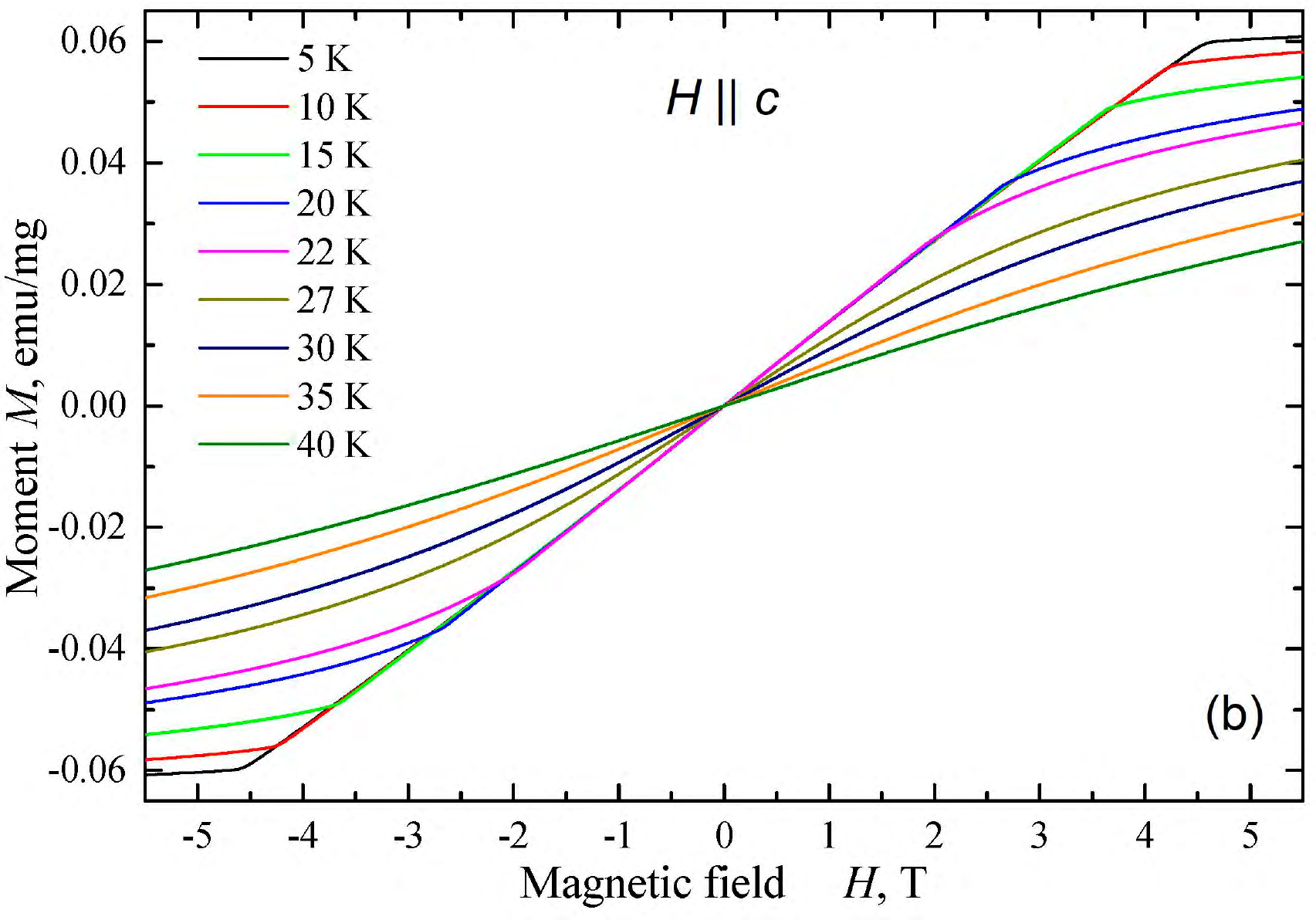}\\
	\includegraphics[width=0.47\columnwidth]{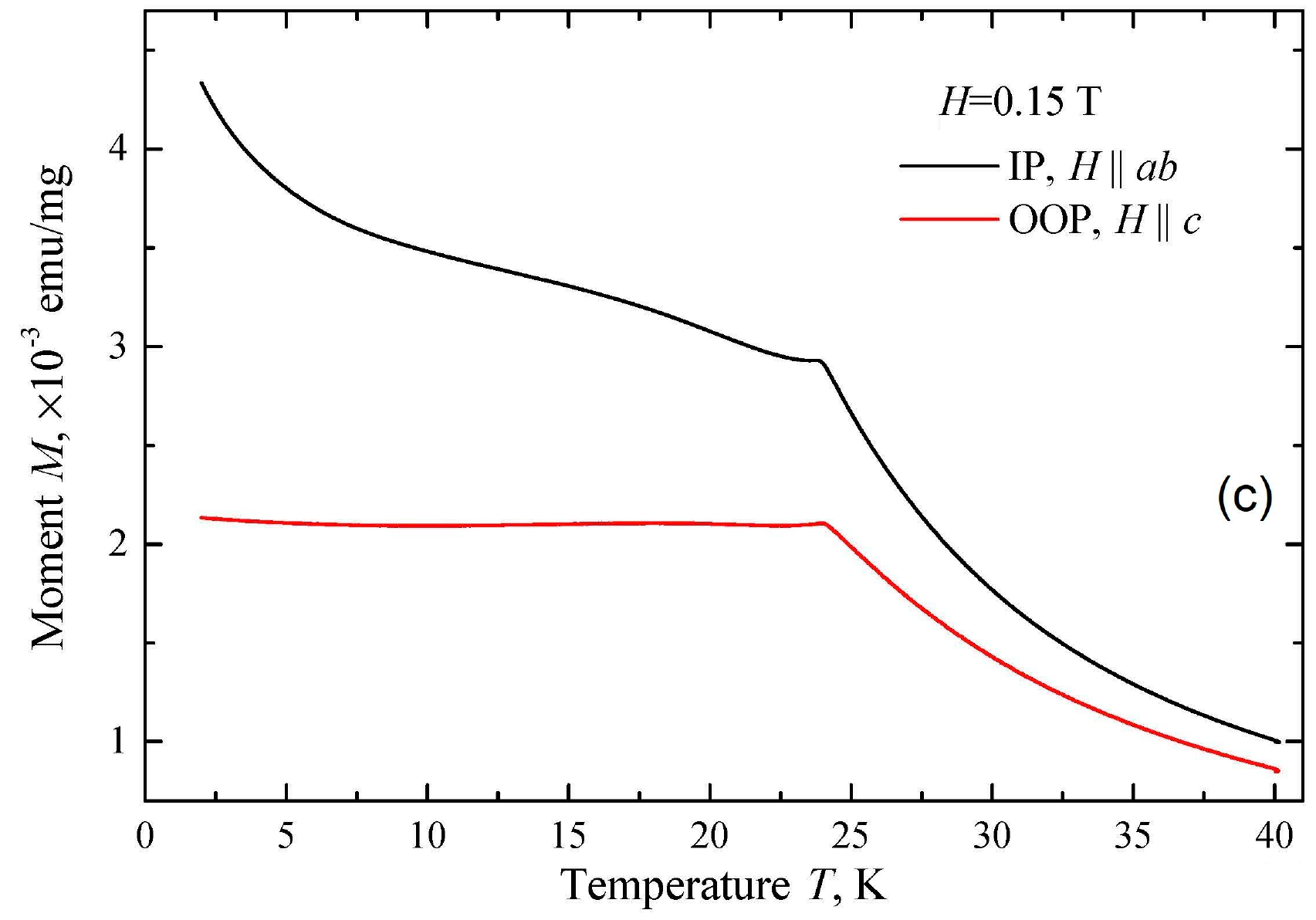}
		\end{center}
{\bf  Fig.\,S1$\: |$ Magnetization dependence on magnetic field  and temperature for EuSn$_2$As$_2$ 
	single crystal $M(H)$ at several fixed temperatures for 	magnetic field applied: a,}  in-plane (IP, along the $ab$ crystal plane), and {\bf b}, out-of-plane (OOP, along the $c$ crystal direction). (c) Temperature dependence of magnetization $M(T)$ measured with DC field  $H= 0.15$\,T for the two orientations of magnetic field. Adapted from Ref.~\cite{golov_JMMM_2022}.\\
		\label{M_exp}
\end{figure*}

\newpage
\section{Supplementary Note 2: 	On the  Topology of ESA Band Structure}

From the band structure (BS) calculations,  Arguilla et al. \cite{arguilla_InChemFront_2017} found Dirac points in the spectrum, essentially above $E_F$. In  subsequent calculations, Li et al. \cite{li_PRX_2019} has suggested EuSn$_2$As$_2$ to be either a strong topological insulator (TI) or axion insulator (AI) \cite{li_PRX_2019, li_PRB_2021}. 

On the experimental side, the first possibility (TI) is inconsistent with the observations as follows:

(i)  monotonic temperature dependence of the zero field resistivity. $\rho(T)$  
decreases smoothly with cooling both in-plane $\rho_{ab}$ and out of plane $\rho_c$,  in the antiferromagnetic and paramagnetic states (except for a minor kink at $T_N$ \cite{chen_ChPhysLet_2020, li_PRB_2021, PRB_tbp}  in the vicinity of the AFM transition, related with magnetic fluctuations). In other words, no signs of surface states (SS) conduction is found in transport;  

(ii) ARPES measurements and  calculations of the electron band structure  don't reveal
 Dirac points, gap, and SS in the close vicinity of the Fermi level.
 
Thus, the experimental results indicate a semimetallic, rather than insulator  nature of the ESA compound with no gap at $E_F$. 

Regarding the existence  of the AI state:  for testing this interesting possibility we measured the  anisotropy of magnetoresistance in the AFM state (for more detail, see Supplementary Note 4). 
Regarding the potential TI phase and the existence of SS states, we note that by time-resolved ARPES measurements \cite{li_PRX_2019} the Dirac crossing point was found to lie 
0.4\,eV above $E_F$ and therefore may not affect charge transport at low temperatures. The absence of Dirac point and of the protected surface states in the vicinity of $E_F$ was also confirmed by  ARPES measurements in Ref.~\cite{PRB_tbp}. 

\newpage
\section{Supplementary Note 3: Alignment of the Eu Magnetic Moments  in ESA}
The type of magnetic ordering in the AFM state of layered vdW semimetals
with magnetic ions (such as EuSn$_2$As$_2$, EuFe$_2$As$_2$, EuSn$_2$P$_2$, EuMg$_2$Bi$_2$,  EuMg$_2$Sb$_2$, etc.)
was debatable in literature several years ago.
There is general agreement in the scientific community 
that the Eu 4f magnetic moments 
in EuSn$_2$As$_2$  and in  other sister compounds 
form an A-type structure in the AFM state. By definition, the A-type structure  means that the $ab$ planes are ferromagnetic and couple antiferromagnetically along the $c$ axis. However,  whether the ferromagnetic moments are directed along the $c$-axis  (the ``easy axis''  configuration), or aligned in the $ab$-plane (``easy-plane'' configuration)  was initially a subject of discussions, which we briefly review below.

In the BS calculations \cite{li_PRX_2019} for ESA it was not determined which of the two configurations, AFM-$b$, or AFM-$c$ has lower energy (and, hence, is realized in practice). In  Figure 1 of Ref.~\cite{li_PRX_2019}, the 
Eu magnetic moments are drawn  along the $c$ axis, though we stress that no evidence for such choice was presented in Ref.~\cite{li_PRX_2019}. Later on, Pakhira et al. \cite{pakhira_PRB_2021} reported  neutron diffraction measurements which also confirmed the Eu  moments are ferromagnetically aligned in the $ab$ plane, {\it lie in the $ab$ plane}, and  are stacked antiferromagnetically along the $c$ axis. 
Strictly speaking, 
the in-plane direction of the FM moments  cannot be determined from neutron diffraction alone. 
For the sister compound, EuFe$_2$As$_2$, from additional symmetry analysis of  magnetic reflections data  in 
neutron diffraction studies \cite{xiao_PRB_2009}  and from magnetic resonant x-ray scattering  \cite{herrero-mart?n_PRB_2009} it was  concluded  that Eu$^{2+}$ moments 
lie  in the ($ab$)-plane. Similar conclusion was drawn for EuMg$_2$Bi$_2$  \cite{pakhira_PRB_2022} by using 
additionally the Bilbao crystallographic server \cite{perez_AnRev Mat_2015}.

This seemingly controversial issue, nevertheless,  may be easily resolved  using  common sence arguments and simple molecular field theory  as follows:

(i) For an isolated layer of Eu atoms with the moments aligned ferromagnetically in-plane, {\em\bf the $ab$-plane alignment is more energetically favorable  than $c$-axis alignment} by $\Delta E = 0.29$\,meV; this corresponds to a magnetic field anisotropy $H_{\rm anis} = \Delta E/(7\mu_B) = 7.25$\,kOe \cite{pakhira_PRB_2021}.

(ii)For the bulk ESA crystal, the measured DC-magnetization $M(H)$ curves (see  Fig.~S1 and also Refs. \cite{golov_JMMM_2022, arguilla_InChemFront_2017, chen_ChPhysLet_2020, pakhira_PRB_2021})  clearly indicate  the saturation field $H_{\rm sf}$ larger for $H\| c$ than for $H\perp c$ field direction. This is because the free energy in the magnetic sublattice expression contains a positive magnetic  anisotropy term $K_U$ which makes the Eu-spins direction $\| (ab)$ preferable 
(see Eq.~(1) of Ref.~\cite{golov_JMMM_2022}).  
Consequently, the saturation fields for in-plane and  out-of-plane orientations are different:
$H_{\rm sf}^{ab} = 2H_e = 2J/ M_s$  
and $H_{\rm sf}^c  = 
2J/ M_s +M_s+2K_U/M_s$, 
where $J$ is the exchange constant, 
$H_e$ is the exchange field, and $M_s$ the saturation magnetization. 
From the measured magnetization curves (Figs.~S1a,b)  at $T \ll T_N$  
we find  $K_U = 1.4\times 10^5$J/m$^3$, the term that {\em\bf ensures  the in-plane Eu-spins alignment  in the AFM state at zero field}. 

(iii) Temperature dependence of the  AC magnetic susceptibility measured in weak fields for  $H\| c$ direction  remains almost constant  below $T=T_N$ at  a value $\chi(T)\approx \chi(T=T_N)$. In contrast, for $H\perp c$  it drops by a factor of two (see  Fig.~8 of Ref.~\cite{pakhira_PRB_2021}). 
This is well consistent with the simple mean field theory  calculations \cite{johnston_PRB_2012, johnston_PRB_2015},   {\em\bf namely for   magnetic moments lying  in the $(ab)$  plane} (see \cite{pakhira_PRB_2021}
and discussion therein). 
It is worth noting, such measurements, to be conclusive, should be done in very low fields ($\leq 100$Gs) and the results are often distorted  by the presence of ferromagnetic impurities and ferromagnetic-type defects in the sample.

{\bf Finally, we note that the debatable issue  of the Eu moments direction in ESA at  zero magnetic field, along the $c$-axis, or along the $(ab)$-plane, is insignificant for the proposed model of NIMR that is applicable for both cases, and is isotropic. Indeed, since $\sigma$ in Eqs. (5) and (6) denotes the spin projection on the AFM order parameter $\boldsymbol{L}_{AFM}$, these and all subsequent formulas remain valid for any direction of $\boldsymbol{L}_{AFM}$.}

\newpage
\section{Supplementary Note 4:  Influence of spin-orbit coupling to the band structure}
The band structure calculated with spin-orbit coupling (SOC) (red lines) is shown in  Figure~S2. There aren't pronounced split bands due to the SOC. General influence of SOC is manifested in the bands shifting. Looking closely, we can see a small splitting of the bands near the Fermi level around ${\Gamma}$-point (insert in the Figure S2) that has a magnitude  about 5~meV. This splitting is missing in case of the SOC absence (blue lines), therefore we associate it with the  spin-orbit interaction. 
In addition, away of the Fermi  energy, there are  areas where the distance between bands has increased in case of SOC included, these are visible 
as small shifts of  about 5--10\,meV. Hence we conclude, the spin-orbit interaction energy can be estimated
no larger than 10~meV  everywhere and $\leq 5$\,meV in the vicinity of $E_F$.
\begin{figure*}[!ht]
	\begin{center}
		\includegraphics[width=0.55\columnwidth]{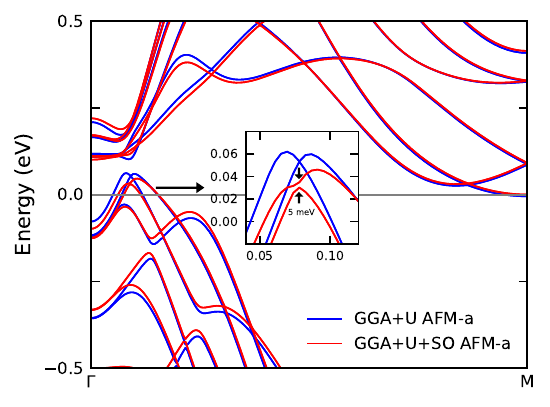}
	\end{center}
	{\bf 
		Fig.\, S2$\: |$ The comparison of GGA+U band structure with (red) and without (blue) spin-orbit interaction in the non-collinear regime (AFM-$a$).}
	\label{dft_band_so}
\end{figure*}

\newpage
\section{Supplementary Note 5:  On the Isotropy of the Negative Magnetoresistance in ESA}

In the band structure calculations for ESA, Li et al. \cite{li_PRX_2019}  considered two  possible configurations: AFM-$b$, and AFM-$c$. It was concluded (i) the PM state in ESA is a strong topological insulator (STI) but ``with no observable gap inside’’, (ii) the AFM-b state in   
ESA is an axion insulator (AI) and topological crystalline insulator  with surface states crossing at E$_F$, and (iii) the AFM-c phase  is AI with gapped SS.

In order to verify the possibility of the predicted AI state, we measured anisotropy of the in-plane magnetoresistance at $T\ll T_N$ by varying the magnetic field $\mathbf{H}$ direction relative to the bias current $\mathbf{J}$ direction. 
Figure S3 shows that the negative magnetoresistance is practically isotropic, within the experimental errors. 
This isotropy  does not support the possibility of the axion insulator state  which is known to be highly anisotropic.

\begin{figure*}[!ht]
	\begin{center}
		\includegraphics[width=0.45\columnwidth]{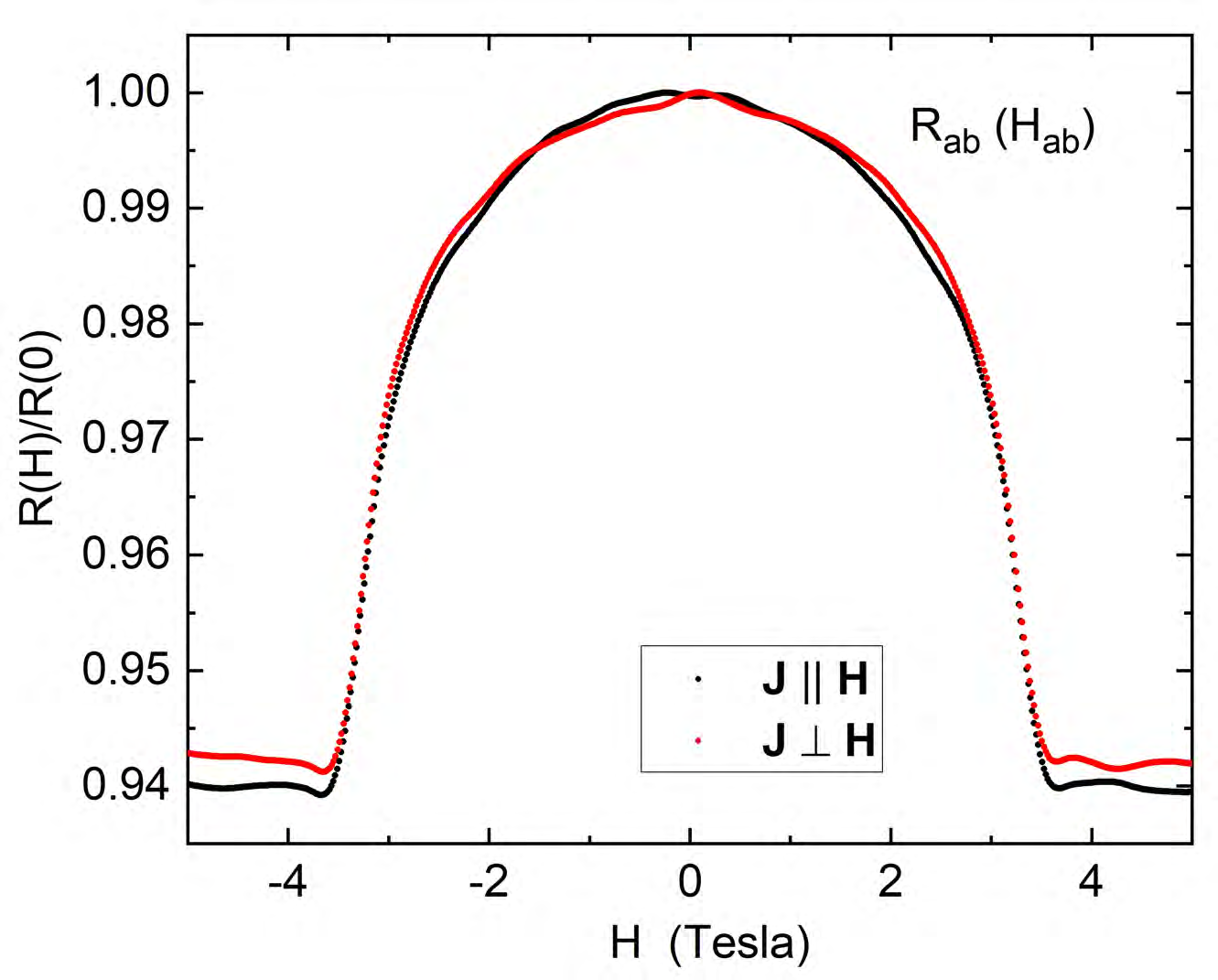}
	\end{center}
	{\bf 
	Fig.\,S3\:$|$ In-plane magnetoresistance $R(H)/R(H=0)$ measured with EuSn$_2$As$_2$ single crystal for magnetic field lying in the $ab$-plane,  and for  two orientations of the field $\mathbf{H}$ relative to the bias current $\mathbf{J}$.}
	\label{anisotropy}
\end{figure*}

\newpage
{\bf Supplementary references}\\